\documentclass{elsart}
\usepackage{color,setspace,times}
\usepackage{amssymb,amsmath,graphicx,color,rotating,subfigure,url,multirow}
\usepackage{lineno}
\usepackage[square,sort&compress,comma]{natbib}

\journal{Physica A} 

\begin{document}

\begin{frontmatter}

\title{Detrended fluctuation analysis of intertrade durations}
\author[BS,SS]{Zhi-Qiang Jiang},
\author[SZSC]{Wei Chen},
\author[BS,SS,RCE,RCSE]{Wei-Xing Zhou\corauthref{cor}}
\corauth[cor]{Corresponding author. Address: 130 Meilong Road, P.O.
Box 114, School of Business, East China University of Science and
Technology, Shanghai 200237, China, Phone: +86 21 64253634, Fax: +86
21 64253152.}
\ead{wxzhou@ecust.edu.cn} %

\address[BS]{School of Business, East China University of Science and Technology, Shanghai 200237, China}
\address[SS]{School of Science, East China University of Science and Technology, Shanghai 200237, China}
\address[SZSC]{Shenzhen Stock Exchange, 5045 Shennan East Road, Shenzhen 518010, China}
\address[RCE]{Research Center for Econophysics, East China University of Science and Technology, Shanghai 200237, China}
\address[RCSE]{Research Center of Systems Engineering, East China University of Science and Technology, Shanghai 200237, China}

\begin{abstract}
The intraday pattern, long memory, and multifractal nature of the
intertrade durations, which are defined as the waiting times between
two consecutive transactions, are investigated based upon the limit
order book data and order flows of 23 liquid Chinese stocks listed
on the Shenzhen Stock Exchange in 2003. An inverse $U$-shaped
intraday pattern in the intertrade durations with an abrupt drop in
the first minute of the afternoon trading is observed. Based on the
detrended fluctuation analysis, we find a crossover of power-law
scaling behaviors for small box sizes (trade numbers in boxes) and
large box sizes and strong evidence in favor of long memory in both
regimes. In addition, the multifractal nature of intertrade
durations in both regimes is confirmed by a multifractal detrended
fluctuation analysis for individual stocks with a few exceptions in
the small-duration regime. The intraday pattern has little influence
on the long memory and multifractaility.
\end{abstract}


\end{frontmatter}

\section{Introduction}

The interevent time, which is defined as the waiting time between
two consecutive events, has attracted considerable interests in
various fields including ecology
\cite{Viswanathan-Afanasyev-Buldyrev-Murphy-Prince-Stanley-1996-Nature,Edwards-Phillips-Watkins-Freeman-Murphy-Afanasyev-Buldyrev-daLuz-Raposo-Stanley-Viswanathan-2007-Nature},
sociology
\cite{Oliveira-Barabasi-2005-Nature,Brokmann-Hufnagel-Geisel-2006-Nature,Barabasi-2005-Nature,Vazquez-2005-PRL,Vazquez-Oliveira-Dezso-Goh-Kondor-Barabasi-2006-PRE},
finance
\cite{Engle-Russell-1998-Em,Engle-Russell-1997-JEF,Engle-2000-Em,Scalas-Gorenflo-Mainardi-2000-PA,Mainardi-Raberto-Gorenflo-Scalas-2000-PA,Masoliver-Montero-Weiss-2003-PRE,Scalas-2006-PA,Masoliver-Montero-Perello-Weiss-2006-JEBO,Jiang-Chen-Zhou-2008-XXX},
seismology
\cite{Telesca-Cuomo-Lapenna-Macchiato-2004-CSF,Corral-2004-PRL,Shcherbakov-Yakovlev-Turcotte-Rundle-2005-PRL,Molchan-2005-PAG,Saichev-Sornette-2006-PRL,Molchan-Kronrod-2007-PAG},
and so on. When we regard the transaction dynamics as a point
process in financial market \cite{Cox-Isham-1980}, trades are
defined as events and intertrade durations are a kind of interevent
time. Several important statistical properties (probability
distribution, long-range dependence and multifractal nature) of the
intertrade durations have been studied.

Empirical analysis of the intertrade durations for different
financial data unveils that the probability distribution can be
described by the Mattag-Leffler function
\cite{Mainardi-Raberto-Gorenflo-Scalas-2000-PA,Sabatelli-Keating-Dudley-Richmond-2002-EPJB},
power laws
\cite{Sabatelli-Keating-Dudley-Richmond-2002-EPJB,Yoon-Choi-Lee-Yum-Kim-2006-PA},
modified power laws
\cite{Masoliver-Montero-Weiss-2003-PRE,Masoliver-Montero-Perello-Weiss-2006-JEBO},
stretched exponentials (or Weibull)
\cite{Bartiromo-2004-PRE,Raberto-Scalas-Mainardi-2002-PA,Ivanov-Yuen-Podobnik-Lee-2004-PRE,Vazquez-Oliveira-Dezso-Goh-Kondor-Barabasi-2006-PRE,Sazuka-2007-PA},
stretched exponentials followed power laws
\cite{Kim-Yoon-2003-Fractals,Kim-Yoon-Kim-Lee-Scalas-2007-JKPS}, to
name a few. Recently, Politi and Scalas rejected the hypothesis that
the waiting time distributions are described by an exponential
\cite{Scalas-Gorenflo-Luckock-Mainardi-Mantelli-Raberto-2004-QF,Scalas-Gorenflo-Luckock-Mainardi-Mantelli-Raberto-2005-FL}
or a power law and found that the $q$-exponential compares well to
the Weibull for waiting time distribution fitting
\cite{Poloti-Scalas-2008-PA}. They also argued that the distribution
differing from an exponential is the consequence of the varying
trade activities during the trading period
\cite{Scalas-Gorenflo-Luckock-Mainardi-Mantelli-Raberto-2004-QF,Scalas-Gorenflo-Luckock-Mainardi-Mantelli-Raberto-2005-FL,Scalas-Kaizoji-Kirchler-Huber-Tedeschi-2006-PA,Politi-Scalas-2007-PA}.
Jiang, Chen, and Zhou found that the intertrade durations exhibit a
scaling behavior and the distribution is Weibull followed by a power
law tail \cite{Jiang-Chen-Zhou-2008-XXX}.

The long memory feature of the intertrade durations is very
important in the ACD model
\cite{Engle-Russell-1998-Em,Engle-Russell-1997-JEF,Engle-2000-Em}
and its variants
\cite{Jasiak-1999-Finance,Bauwens-Giot-2000-AES,Zhang-Russell-Tsay-2001-JEm,Bauwens-Veredas-2004-JEm}.
In the econophysics community, to our best knowledge, the first
research was conducted by Ivanov {\em{et al.}}, applying the
detrended fluctuation analysis (DFA) approach to analyze the
intertrade durations of 30 stocks listed on the NYSE from January
1993 to December 1996 \cite{Ivanov-Yuen-Podobnik-Lee-2004-PRE}. They
found that there are two scaling ranges in the fluctuation function,
where long-range power law correlations within a trading day
followed by a crossover to even stronger correlations over time
scale more than one trading day. They also argued that the
appearance of two scaling regimes is linked to the timescales over
which information disseminates. Yuen and Ivanov further analyzed the
intertrade times of 100 stocks listed on the NYSE and 100 stocks
traded on the NASDAQ and found that the crossover behavior also
exists for all stocks on both markets \cite{Yuen-Ivanov-2005-XXX}.
In addition, it is found that the stocks on NASDAQ show much
stronger correlations within one trading day than that on the NYSE,
albeit both markets display the same memory feature for time scale
larger than one day. This result is interpreted by the institutional
difference between the two markets (multi-dealer in NASDAQ, one
market maker in NYSE). By investigating 3924 stocks from 1994 to
1995 and 4044 stocks in the whole year 2000 traded on the NYSE,
Eisler and Kert{\'e}sz found the crossover behavior again based on
the fluctuation analysis \cite{Eisler-Kertesz-2006-EPJB}. More
interesting, the Hurst exponents of the intertrade durations
decrease with the logarithm of mean intertrade duration, $H = H^* -
\gamma_T \log\langle T \rangle$.

In addition, the multifractal nature in the intertrade durations of
30 DAX stocks (from 28 November 1997 to 31 December 1999)  was
studied by O{\'s}wi{\c{e}}cimka {\em{et al.}}, based on the
multifractal detrended fluctuation analysis (MFDFA) approach
\cite{Oswiecimka-Kwapien-Drozdz-2005-PA}. This property has not been
well documented for other markets. In this work, we shall perform
detailed (multifractal) detrended fluctuation analyses of the
intertrade durations of 23 Chinese stocks traded on the Shenzhen
Stock Exchange in 2003. We find that the intraday pattern of the
intertrade durations does not have significant impact on the
long-range dependence and the multifractal nature. We note that the
multifractal nature of the returns, the capital fluxes, and the
bid-ask spreads of Chinese stocks has been investigated thoroughly
\cite{Gu-Chen-Zhou-2007-EPJB,Jiang-Guo-Zhou-2007-EPJB,Jiang-Zhou-2007-PA,Jiang-Zhou-2008a-PA,Jiang-Zhou-2008b-PA,Jiang-Zhou-2008-XXX}.
The current work thus complements this literature.

This paper is organized as follows. In Section \ref{S1:Data}, we
briefly describe the data sets adopted. Section
\ref{S1:intradaypattern} investigates the intraday pattern of
intertrade durations. Section \ref{S1:DFA} studies the memory
behavior and the multifractal nature of the intertrade durations
using the DFA approach. Section \ref{S1:Conclusion} concludes.

\section{Data sets}
\label{S1:Data}

The Chinese stock market is an order-driven market. The organized
stock market in mainland China is composed of two stock exchanges,
the Shenzhen Stock Exchange (SZSE) and the Shanghai Stock Exchange
(SHZE). On the SZSE, each trading day is partitioned into three
parts before 1 July 2007, named open call action, cooling period,
and continuous double auction. The open call action begins at 9:15
AM and ends at 9:25 AM. Orders are allowed to be submitted and
canceled before 9:20 AM. After 9:20, order cancelation is
prohibited. At 9:25 AM, part of the submitted orders are executed
based on the maximal transaction volume principle, while unsatisfied
orders are left on the order book. It is followed by a cooling
period from 9:25 AM to 9:30 AM. During the cooling period, all
orders are allowed to add into the limit-order book, but no one is
executed till 9:30 AM. The continuous double auction operates from
9:30 AM to 11:30 AM and from 13:00 PM to 15:00 PM. According to
price-time priority, transaction occurs based on one by one matching
of incoming effective market orders and limit orders waiting on the
limit-order book. Note that the time interval from 11:30 AM to 13:00
PM is also a cooling period for lunch. Our primary purpose is to
investigate the waiting time between two consecutive transactions.
Hence, only trades during the continuous auction are considered in
this work.

Our study is based on the data of the limit-order books of 23 liquid
stocks listed on the Shenzhen Stock Exchange (SZSE) in the whole
year 2003. These stocks are representative since they were included
as constituents in the Shenzhen Component Index. The limit-order
book records ultra-high-frequency data whose time stamps are
accurate to 0.01 second including details of every event. Assuming
that there are $n$ trades at times $\{t_i: i = 1, 2, \cdots, n\}$
during the time interval from 9:30 AM to 11:30 AM or from 13:00 PM
to 15:00 PM on a trading day, we obtain $n-1$ intertrade durations
$\tau_i = t_{i+1} -t_i$ with $i=1,2,\cdots,n-1$. The variables of
time are in units of second. In addition, we stress that no
intertrade duration is calculated between two trades overnight or
crossing the noon closing. Although the time resolution of our data
is as precise as 0.01 second, there are still trades stamped with
the same time, indicating that the intertrade duration is vanishing
between the two corresponding trades. For convenience, we treat the
trades occurring at the same time as one trade at that time.
Therefore, vanishing durations are excluded. For the 23 stocks, the
average intertrade duration varies from 3.8 seconds to 49.4 seconds
\cite{Jiang-Chen-Zhou-2008-XXX}.

\section{Intraday pattern}
\label{S1:intradaypattern}

Many empirical studies show that the high-frequency financial
variables exhibit intraday patterns, such as the returns
\cite{Cornell-Schwarz-Szakmary-1995-JBanF}, volatilities
\cite{Andersen-Bollerslev-1997-JEF}, bid-ask spreads
\cite{Mcinish-Wood-1992-JF,Chan-Christie-Schultz-1995-JB,Gu-Chen-Zhou-2007-EPJB,Ni-Zhou-2007-XXX},
trading volumes
\cite{Jain-Joh-1988-JFQA,Admati-Pfleiderer-1988-RFS}, and so on. The
intertrade durations are also found to exhibit an inverse $U$-shaped
pattern in the NYSE market
\cite{Engle-Russell-1998-Em,Hafner-2005-QF} and Russian stock market
\cite{Anatolyev-Shakin-2007-AFE}, which indicates higher trading
activities in the open and close than in other time during each
trading day. Therefore, it is necessary to investigate the intraday
patterns in the intertrade durations of the Chinese stocks under
investigation to check if such patterns have influence on the
possible long-range dependence and multifractal property.

We segment the continuous double auction of each trading day into
240 successive 1-min intervals. For a given stock, we define an
average intertrade duration for each interval as follows,
\begin{equation}
 \tau_{ij} = \frac{1}{N_{ij}} \sum_{k=1}^{N_{ij}} \tau_k,
  \label{Eq:tau:ij}
\end{equation}
where $\tau_{ij}$ is the average duration of the $j$-th interval in
the $i$-th trading day, $N_{ij}$ represents the number of intertrade
durations of the $j$-th interval in the $i$-th trading day. The
average intertrade duration in the $j$-th time interval is
calculated as follows,
\begin{equation}
 \langle \tau \rangle_j = \frac{1}{N_d} \sum_{i=1}^{N_d} \tau_{ij},
  \label{Eq:meanduration}
\end{equation}
where $N_d$ is the number of trading days.

\begin{figure}[htb]
\centering
\includegraphics[width=6.5cm]{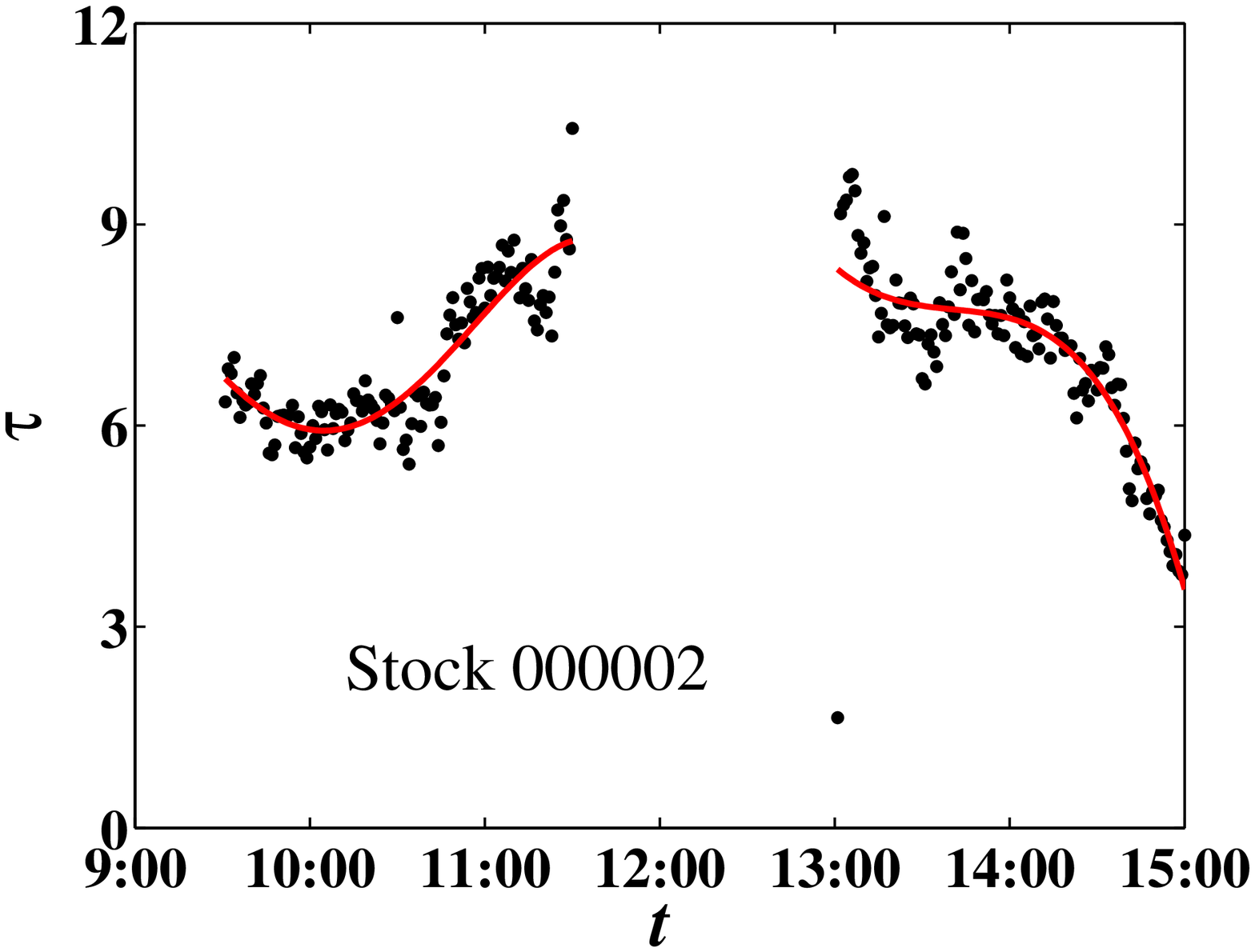}
\includegraphics[width=6.5cm]{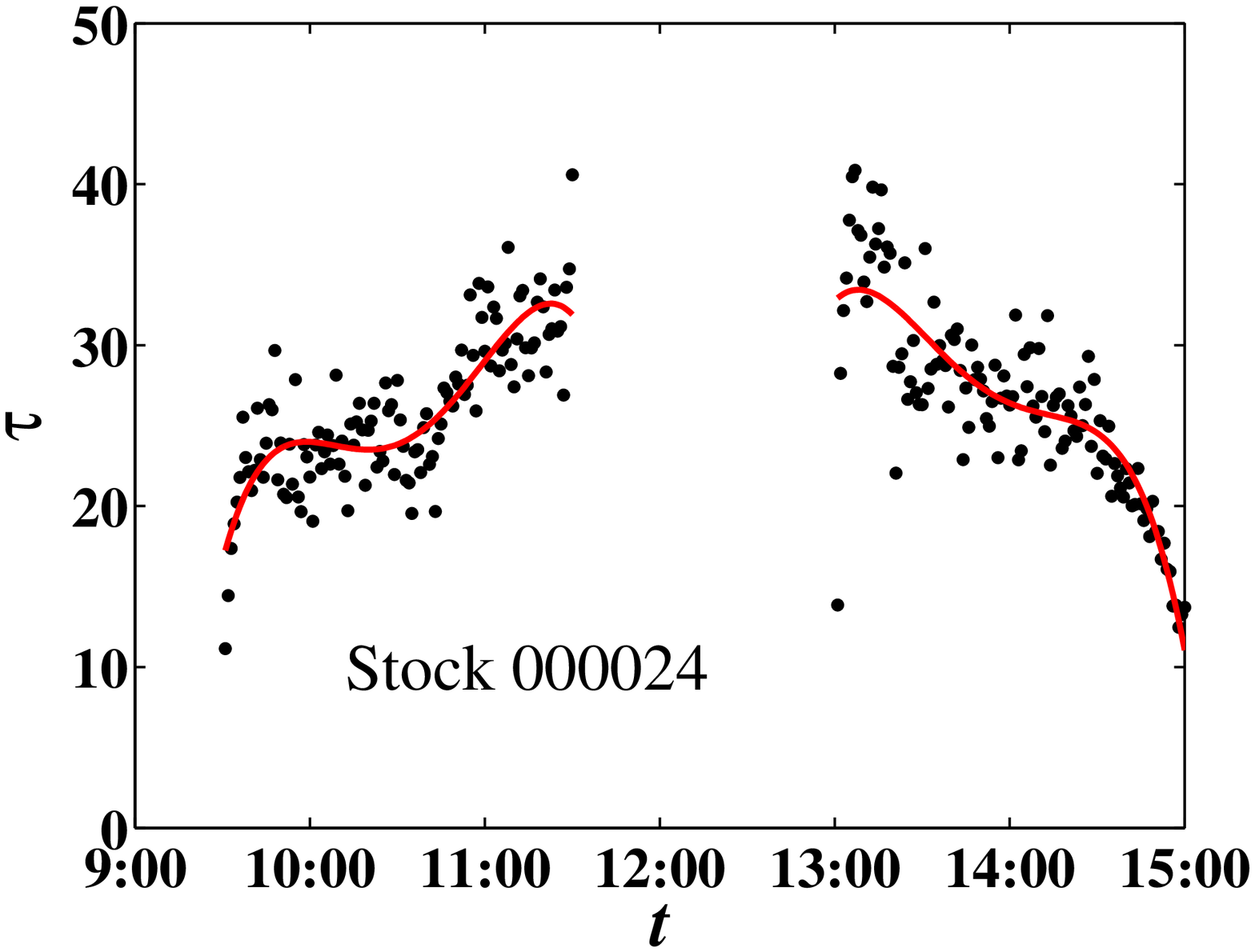}
\includegraphics[width=6.5cm]{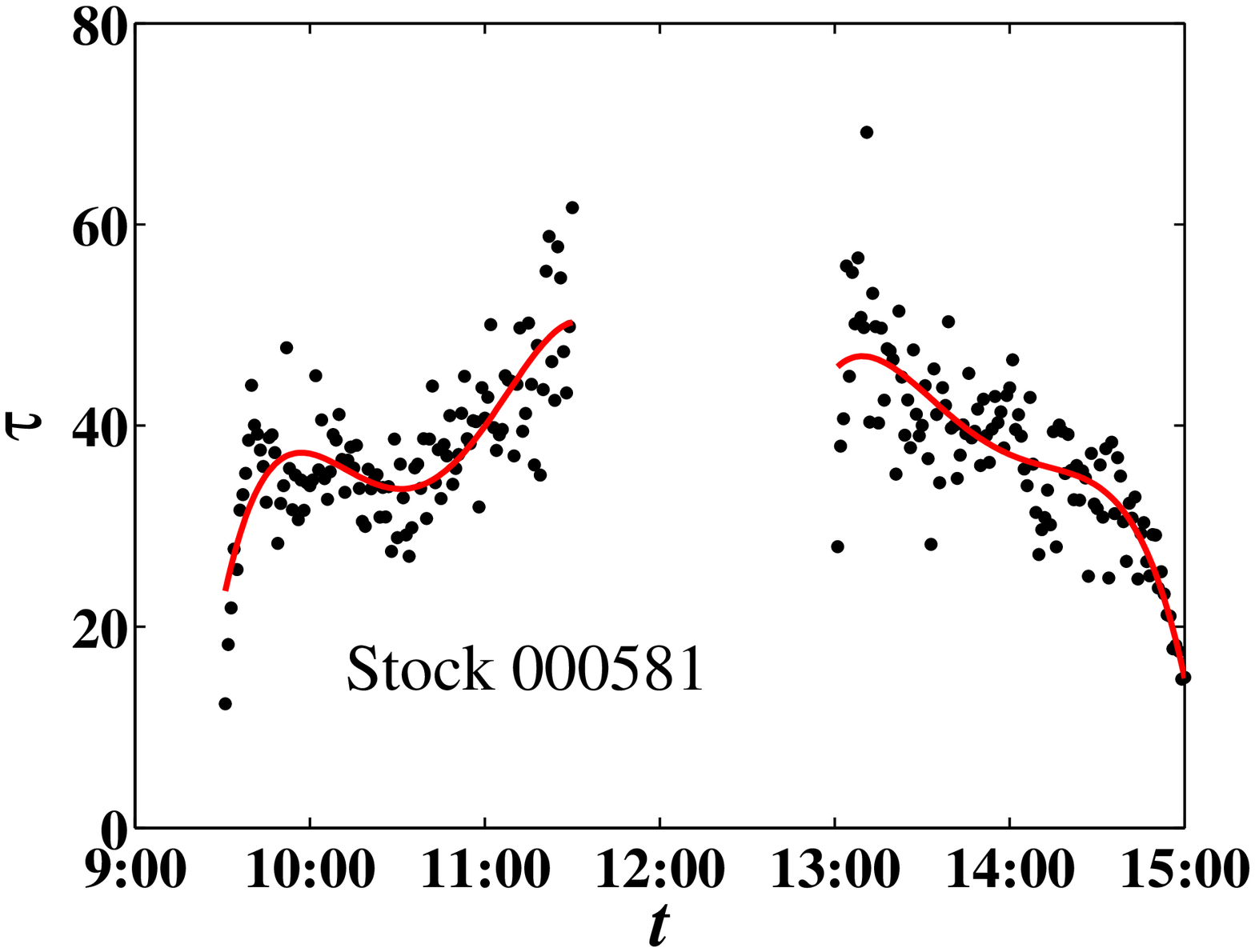}
\includegraphics[width=6.5cm]{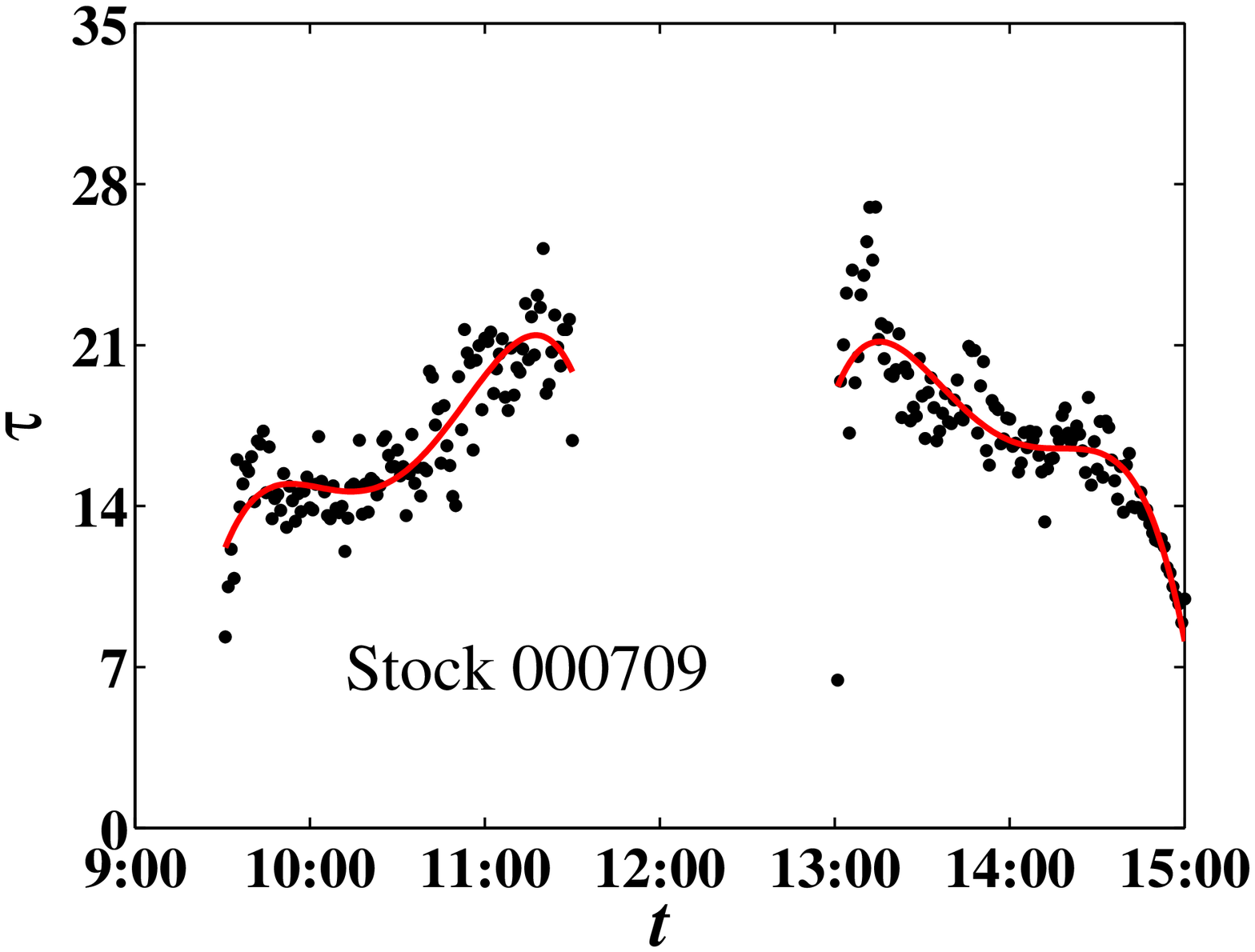}
\caption{(Color online) Intraday pattern of intertrade durations for
four different stocks traded on the SZSE during the calendar year
2003. The full circles are the average intertrade durations. The
continuous curves are the polynomial fits to the data.}
\label{Fig:IntradayPattern}
\end{figure}

Four stocks (000002, 000024, 000581, 000709) are randomly chosen
from the 23 stocks as typical examples to illustrate the results.
Fig.~\ref{Fig:IntradayPattern} depicts the intraday pattern of the
intertrade durations for the four stocks. The durations exhibit a
crude inverse $U$-shaped pattern. For most stocks, the mean
durations in the open and close are much smaller than those in the
rest time of the trading day, which indicates heavier trades in the
open and close. Our results are in line with the results of IBM
transaction data \cite{Engle-Russell-1998-Em}. More interesting, the
duration during the first minute in the afternoon is very low. This
phenomenon arises from the institutional features of the Chinese
stock market. The traders can submit orders during the nontrading
period from 11:30 to 13:00 and these orders are disposed immediately
at 13:00, which leads to very high trading activity and small
average duration during the 1-min interval right after 13:00.

\section{Detrended fluctuation analysis}
\label{S1:DFA}

\subsection{Long-range dependence}

We first study the temporal correlation in the intertrade duration
series. The detrended fluctuation analysis is utilized, which has
the ability to extract long-range power-law correlation in
non-stationary time series
\cite{Peng-Buldyrev-Havlin-Simons-Stanley-Goldberger-1994-PRE,Kantelhardt-Bunde-Rego-Havlin-Bunde-2001-PA}.
For a given intertrade duration series $\{\tau_i|i =
1,2,\cdots,N\}$, we can define the cumulative summation series $y_i$
as follows,
\begin{equation}
 y_i = \sum_{j=1}^{i} \tau_j,~~i = 1,2,\cdots,N.
  \label{Eq:cumsum}
\end{equation}
The series $y$ is covered by $N_s$ disjoint boxes with the same size
$s$. When the whole series $y_i$ cannot be completely covered by
$N_s$ boxes, we can utilize $2N_s$ boxes to cover the series from
both ends of the series. In each box, a cubic polynomial trend
function $g$ of the sub-series is determined. The local detrended
fluctuation function $f_k(s)$ in the $k$-th box is defined as the
r.m.s. of the fitting residuals:
\begin{equation}
 [f_k(s)]^2 = \frac{1}{s}\sum_{i=(k-1)s+1}^{ks} [y_i-g_i]^2~.
  \label{Eq:fk:s}
\end{equation}
The overall detrended fluctuation is estimated as follows
\begin{equation}
 [F_2(s)]^2 = \frac{1}{N_s}\sum_{i=1}^{N_s} \left[f_k(s)\right]^2.
  \label{Eq:F2:s}
\end{equation}
As the box size $s$ varies in the range of $[20,N/4]$, one can
determine a power law relationship between the overall fluctuation
function $F_2(s)$ and the box size $s$, which reads,
\begin{equation}
 F_2(s) \sim s^H,
  \label{Eq:Hurst}
\end{equation}
where $H$ signifies the Hurst index, which is related to the power
spectrum exponent $\eta$ by $\eta = 2H-1$ and to the autocorrelation
exponent $\gamma$ by $\gamma = 2-2H$.

\begin{figure}[htb]
\centering
\includegraphics[width=6.5cm]{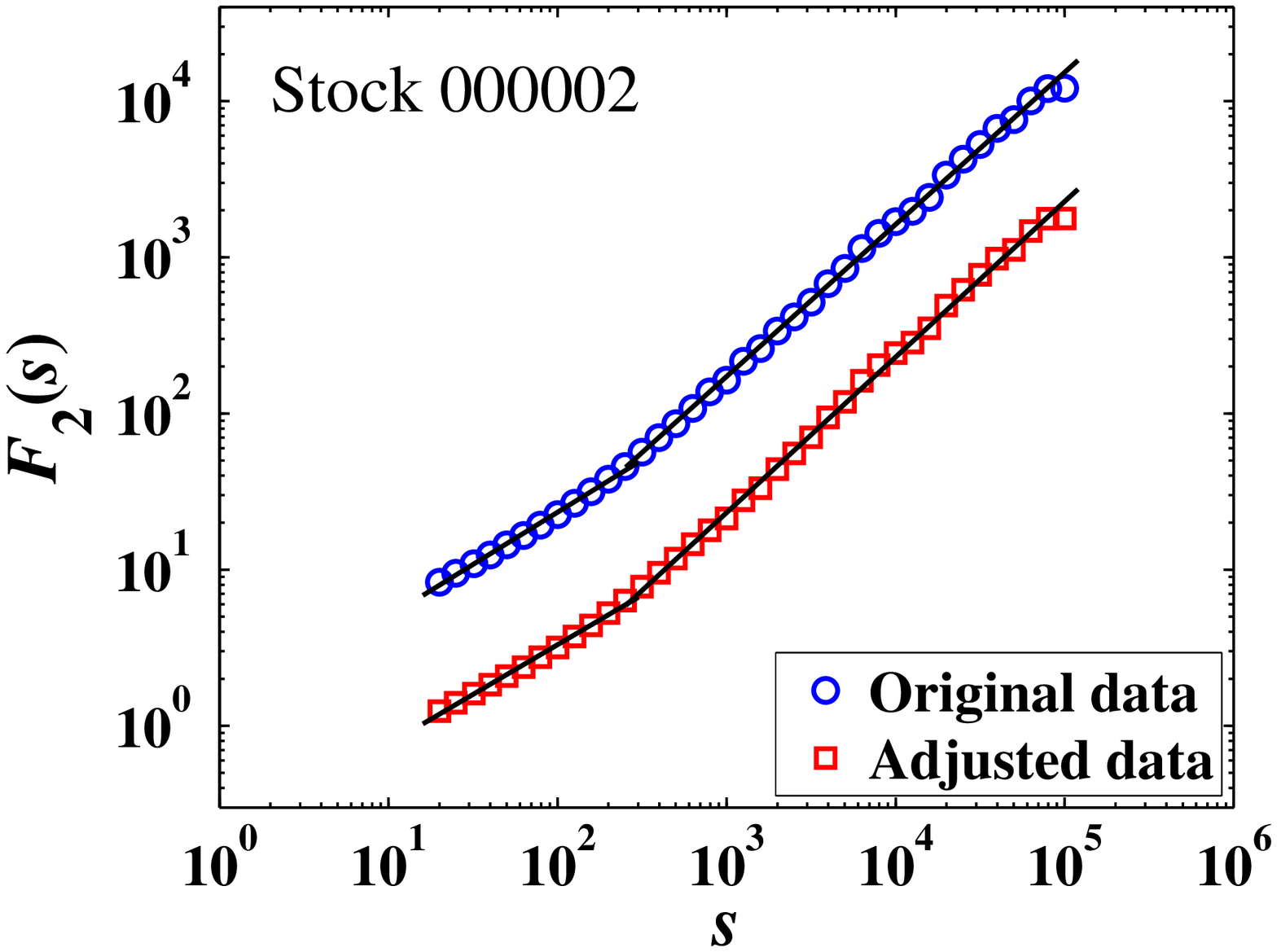}
\includegraphics[width=6.5cm]{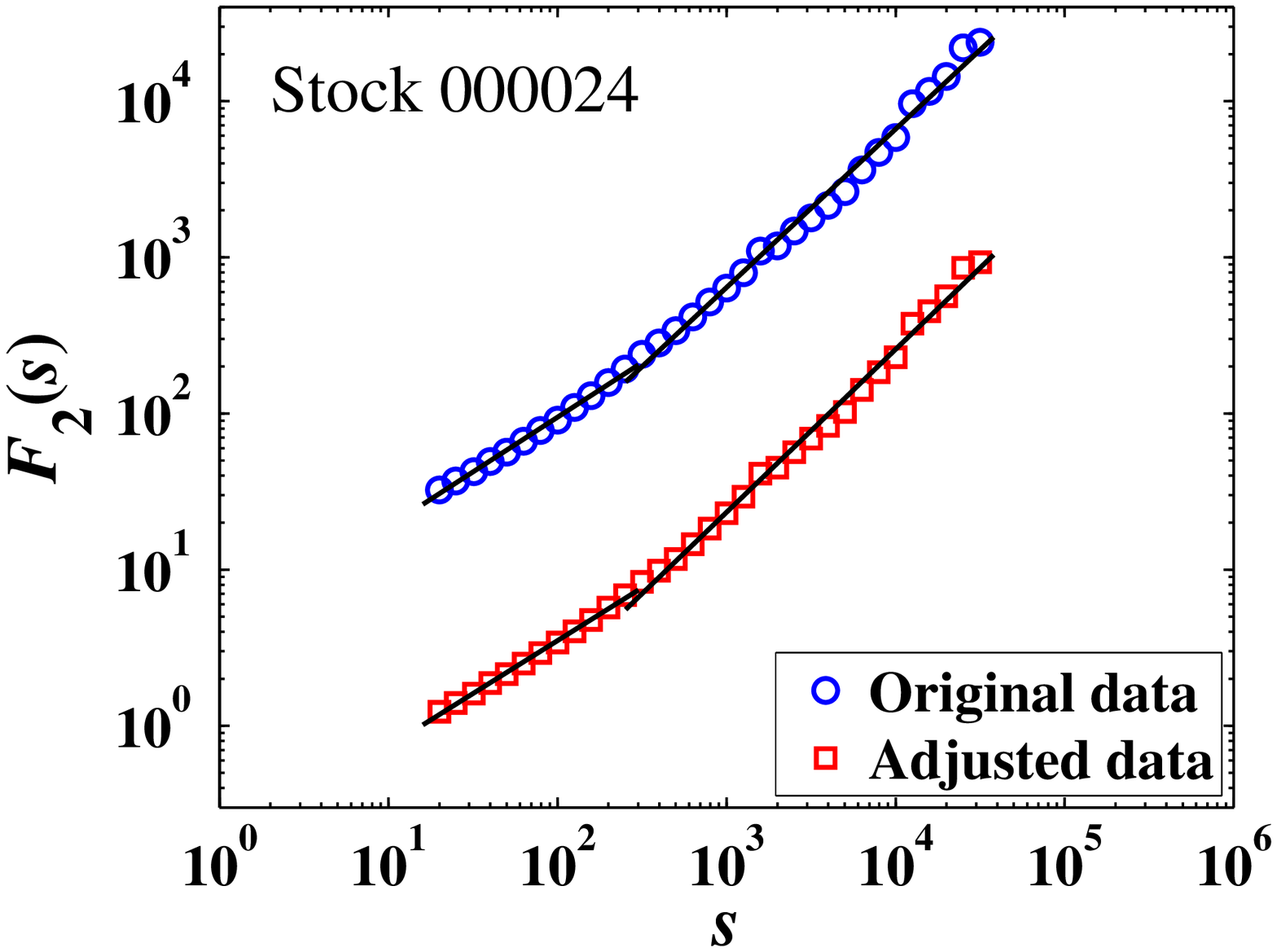}
\includegraphics[width=6.5cm]{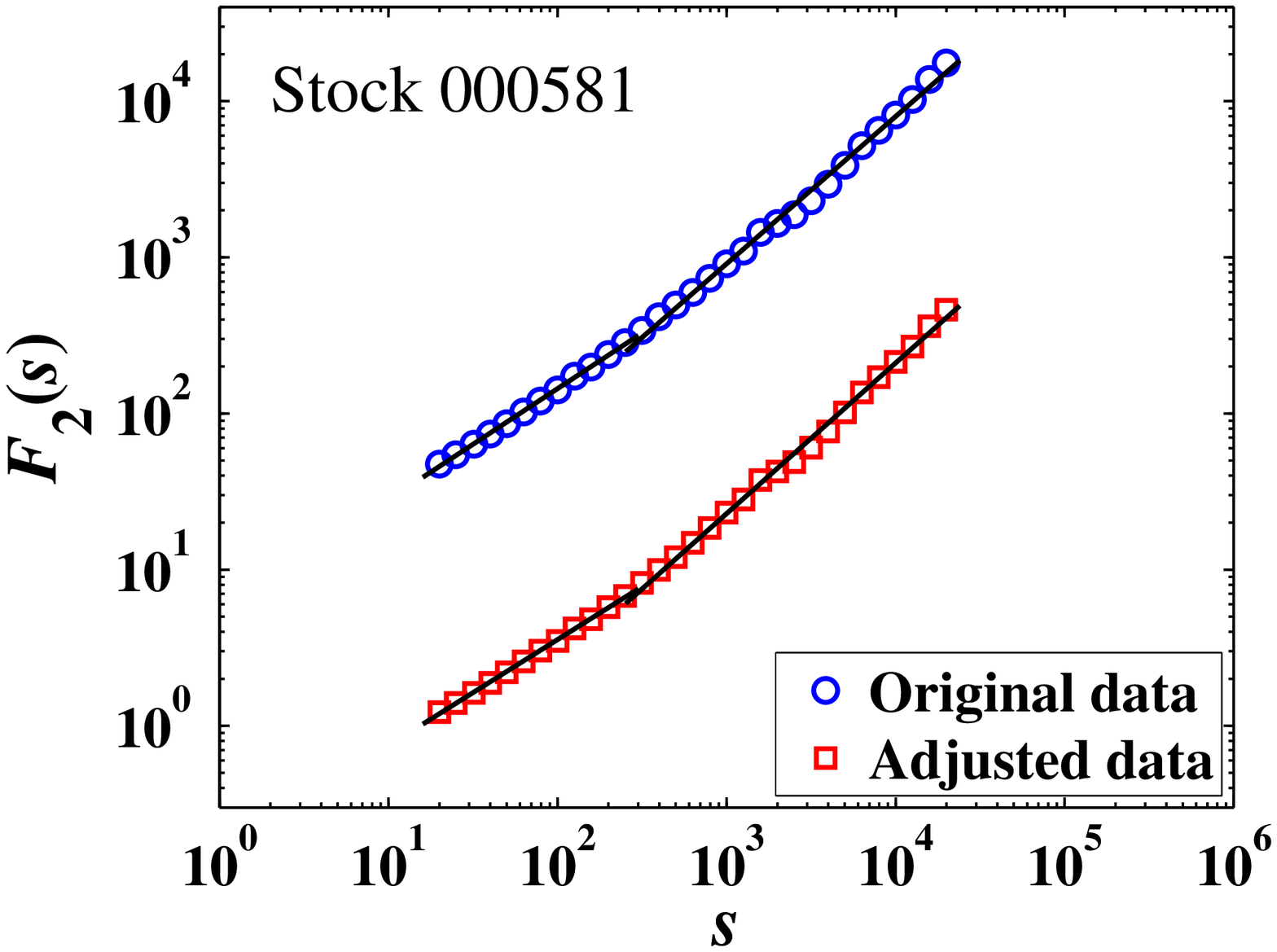}
\includegraphics[width=6.5cm]{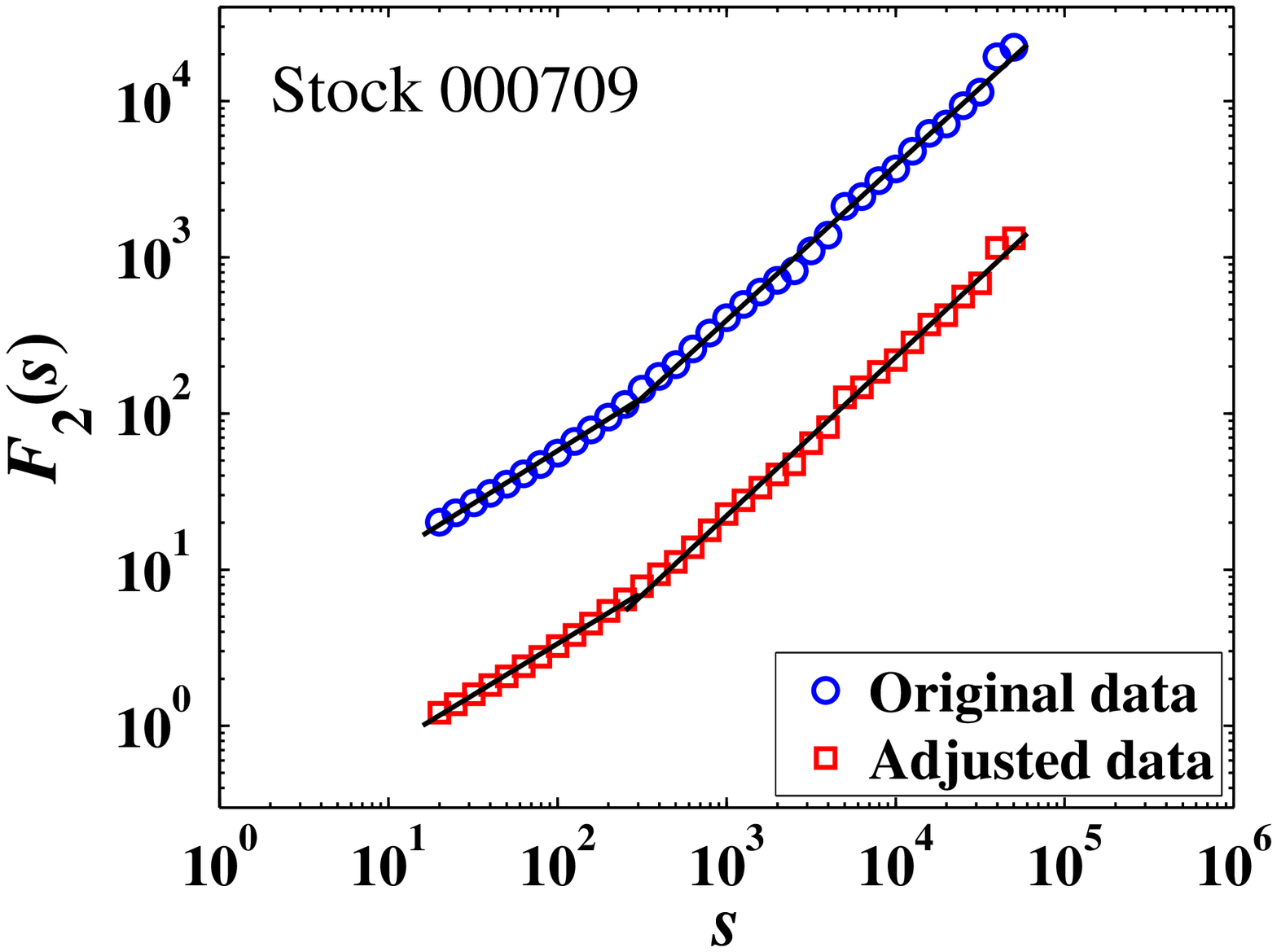}
\caption{(Color online) Log-log plots of the overall fluctuation
function $F_2(s)$ with respect to the box size $s$ for four
different stocks with $q = 2$. The open circles and squares stand
for the original and adjusted data, respectively. The solid lines
are the best fits to the data in the scaling ranges.}
\label{Fig:DFALM}
\end{figure}

We apply the DFA approach to analyze both the original data and the
adjusted data. The adjusted data are obtained by removing the
intraday pattern from the original data:
\begin{equation}
 \widetilde{\tau} = {\tau_t}/{\langle \tau \rangle_j}~,
  \label{Eq:RIP}
\end{equation}
where $\langle \tau \rangle_j$ is the average duration in the $j$-th
1-min interval to which the $t$-th trade belongs. Performing the
same analysis on adjusted data can test whether the long dependence
of intertrade duration stems from the seasonal periodicity.
Fig.~\ref{Fig:DFALM} shows the log-log plots of the overall
fluctuations $F_2(s)$ as a function of the box size $s$ for the
selected four stocks. For each stock, one can observe a crossover
from a scaling range with a lower exponent $H_1$ over $s <
s_{\times}$ to a scaling range with a higher exponent $H_2$ over $s
> s_{\times}$ in the scaling curves for both original and adjusted
data. Note that the crossover happens at about $s_{\times} \approx
300$. We find that there is only one stock (000720) which does not
have a crossover of regimes and, as shown in
Table~\ref{Tb:HurstIndex}, $H_1=H_2$ for this stock. Indeed, this
stock exhibited very different behavior, whose prices were
controlled and manipulated by block investors \cite{Zhou-2007-XXX}.

The Hurst indexes $H_1$ and $H_2$ for all the stocks are listed in
Table~\ref{Tb:HurstIndex}. One can see that both $H_1$ and $H_2$ are
significantly greater than 0.5, indicating the long-range dependence
in the intertrade durations. Except for stock 000720, we observe
that $H_2
> H_1$ for all the 22 remaining stocks, which means much stronger
correlation for larger $s$ in the intertrade durations. Excluding
stock 000720, we find that $\langle H_1 \rangle = 0.69 \pm 0.02$ and
$\langle H_2 \rangle = 0.96 \pm 0.04$ for the original data and
$\langle H_1 \rangle = 0.67 \pm 0.02$ and $\langle H_2 \rangle =
0.98 \pm 0.04$ for the adjusted data. We can conclude that the
intraday pattern has minor influence on the long-range dependence in
the durations. Furthermore, a systematic difference between the
scaling exponents $\langle H_2 - H_1 \rangle$ is $0.26 \pm 0.6$ for
the original data and $0.31 \pm 0.06$ for the adjusted data, which
is comparable to the value $0.30 \pm 0.05$ for the NYSE
\cite{Ivanov-Yuen-Podobnik-Lee-2004-PRE}. In addition, stocks with
higher trade activities appear to have stronger autocorrelations in
the NYSE and NASDAQ
\cite{Eisler-Kertesz-2006-EPJB,Yuen-Ivanov-2005-XXX}. However, in
our dataset, we find no clear dependence between the Hurst index and
the average trading activity measured by mean intertrade duration.
This observation is far from conclusive, since our database contains
only 23 stocks.

\begin{table}[htb]
\begin{center}
\caption{\label{Tb:HurstIndex}Hurst indexes $H_1$ and $H_2$, mean
trades per day $\langle N_T \rangle$, and the width of multifractal
spectrum $\Delta \alpha$ of the intertrade durations for the 23
stocks.}
\medskip
\begin{tabular}{cccccccccc}
  \hline\hline
  \multirow{3}*[3.6mm]{Code} & \multirow{3}*[3.6mm]{$\langle N_T \rangle$}&& \multicolumn{3}{c}{Original data}&&\multicolumn{3}{c}{Adjusted data} \\  %
  \cline{4-6}  \cline{8-10}
         &&& $H_1$ & $H_2$ & $\Delta \alpha$ && $H_1$ & $H_2$ & $\Delta \alpha$  \\\hline
  000001 & 3676 && $0.65 \pm 0.01$ & $0.97 \pm 0.01$ & 0.91 && $0.61 \pm 0.01$ & $0.99 \pm 0.01$ & 0.81\\
  000002 & 2084 && $0.67 \pm 0.01$ & $0.97 \pm 0.01$ & 0.90 && $0.64 \pm 0.01$ & $1.00 \pm 0.01$ & 0.82\\
  000009 & 1842 && $0.68 \pm 0.01$ & $1.01 \pm 0.01$ & 0.97 && $0.65 \pm 0.02$ & $1.03 \pm 0.01$ & 0.96\\
  000012 & 1210 && $0.71 \pm 0.01$ & $0.94 \pm 0.01$ & 0.60 && $0.69 \pm 0.01$ & $0.96 \pm 0.01$ & 0.59\\
  000016 & 778  && $0.68 \pm 0.01$ & $0.97 \pm 0.01$ & 0.83 && $0.66 \pm 0.01$ & $1.00 \pm 0.01$ & 0.79\\
  000021 & 1695 && $0.70 \pm 0.02$ & $0.97 \pm 0.02$ & 0.83 && $0.68 \pm 0.02$ & $0.99 \pm 0.01$ & 0.77\\
  000024 & 553  && $0.70 \pm 0.01$ & $1.01 \pm 0.01$ & 0.82 && $0.68 \pm 0.01$ & $1.04 \pm 0.02$ & 0.76\\
  000027 & 1275 && $0.69 \pm 0.01$ & $0.97 \pm 0.01$ & 0.66 && $0.68 \pm 0.01$ & $0.99 \pm 0.01$ & 0.64\\
  000063 & 1073 && $0.72 \pm 0.01$ & $0.95 \pm 0.01$ & 0.76 && $0.71 \pm 0.02$ & $0.96 \pm 0.02$ & 0.76\\
  000066 & 1146 && $0.69 \pm 0.01$ & $0.94 \pm 0.01$ & 0.77 && $0.68 \pm 0.01$ & $0.96 \pm 0.01$ & 0.78\\
  000088 & 376  && $0.70 \pm 0.01$ & $0.95 \pm 0.01$ & 0.68 && $0.68 \pm 0.01$ & $0.96 \pm 0.01$ & 0.67\\
  000089 & 775  && $0.68 \pm 0.01$ & $1.01 \pm 0.01$ & 0.85 && $0.67 \pm 0.01$ & $1.02 \pm 0.01$ & 0.79\\
  000406 & 1116 && $0.69 \pm 0.01$ & $0.99 \pm 0.01$ & 0.90 && $0.67 \pm 0.01$ & $1.01 \pm 0.01$ & 0.81\\
  000429 & 488  && $0.69 \pm 0.01$ & $0.91 \pm 0.01$ & 1.03 && $0.67 \pm 0.01$ & $0.93 \pm 0.01$ & 0.98\\
  000488 & 496  && $0.70 \pm 0.01$ & $0.91 \pm 0.01$ & 0.77 && $0.68 \pm 0.01$ & $0.94 \pm 0.01$ & 0.73\\
  000539 & 417  && $0.73 \pm 0.01$ & $0.82 \pm 0.01$ & 1.00 && $0.69 \pm 0.01$ & $0.85 \pm 0.01$ & 1.01\\
  000541 & 283  && $0.69 \pm 0.01$ & $0.92 \pm 0.01$ & 0.87 && $0.67 \pm 0.01$ & $0.95 \pm 0.01$ & 0.82\\
  000550 & 1405 && $0.71 \pm 0.02$ & $0.94 \pm 0.02$ & 0.76 && $0.70 \pm 0.02$ & $0.95 \pm 0.01$ & 0.69\\
  000581 & 373  && $0.71 \pm 0.01$ & $0.94 \pm 0.01$ & 0.78 && $0.68 \pm 0.01$ & $0.96 \pm 0.01$ & 0.78\\
  000625 & 1643 && $0.70 \pm 0.01$ & $0.97 \pm 0.01$ & 0.63 && $0.69 \pm 0.01$ & $0.99 \pm 0.01$ & 0.52\\
  000709 & 853  && $0.68 \pm 0.01$ & $0.99 \pm 0.01$ & 0.69 && $0.66 \pm 0.01$ & $1.02 \pm 0.01$ & 0.61\\
  000720 & 486  && $0.85 \pm 0.01$ & $0.86 \pm 0.01$ & 0.61 && $0.83 \pm 0.01$ & $0.88 \pm 0.01$ & 0.59\\
  000778 & 651  && $0.69 \pm 0.01$ & $1.02 \pm 0.01$ & 0.72 && $0.68 \pm 0.01$ & $1.05 \pm 0.01$ & 0.64\\
  \hline\hline
\end{tabular}
\end{center}
\end{table}

\subsection{Multifractal nature}

In this section, we apply a multifractal detrended fluctuation
analysis to investigate the multifractal nature of intertrade
durations. The overall detrended fluctuation in Eq.~(\ref{Eq:F2:s})
is generalized to the following form
\begin{equation}
 F_q(s) = \left\{\frac{1}{N_s}\sum_{k=1}^{N_s}[f_k(s)]^q
 \right\}^{1/q}~,
  \label{Eq:Fq}
\end{equation}
where $q$ can take any real number except $q = 0$. When $q = 0$, we
have
\begin{equation}
 F_0(s) = \exp\left\{\frac{1}{N_s}\sum_{k=1}^{N_s}\ln[f_k(s)] \right\}.
  \label{Eq:F0}
\end{equation}
By varying the value of $s$ in the range from $s_{\min} = 20$ to
$s_{\max} = N/4$, one can expect the detrended fluctuation function
$F_q(s)$ scales with the size $s$:
\begin{equation}
 F_q(s) \sim s^{h(q)},
  \label{Eq:scaling}
\end{equation}
where $h(q)$ is the generalized Hurst index. Note that when $q = 2$,
$h(2)$ is nothing but the Hurst index $H$. The scaling exponent
function $\tau(q)$, which is used to reveal the multifractality in
the standard multifractal formalism based on partition function, can
be obtained numerically as follows:
\begin{equation}
 \tau(q) = qh(q) - D_f,
  \label{Eq:scalingfunction}
\end{equation}
where $D_f$ is the fractal dimension of the geometric support of the
multifractal measure (in the current case $D_f = 1$). The local
singularity exponent $\alpha$ and its spectrum $f(\alpha)$ are
related to $\tau(q)$ through the Legendre transformation
\cite{Halsey-Jensen-Kadanoff-Procaccia-Shraiman-1986-PRA},
\begin{equation}
\left\{ \begin{aligned}
         \alpha &= {\rm{d}}\tau(q)/{\rm{d}}q \\
                  f(\alpha)&=q \alpha -\tau(q)
        \end{aligned} \right.~.
\label{Eq:alphaf}
\end{equation}
Since the size of each time series is finite, the estimate of
$F_q(s)$ will fluctuate remarkably for large values of $|q|$,
especially for large $s$. We focus on $q \in [-4, 6]$ to obtain
reasonable statistics in the estimation of $F_q(s)$.

\begin{figure}[htb]
\centering
\includegraphics[width=3.3cm]{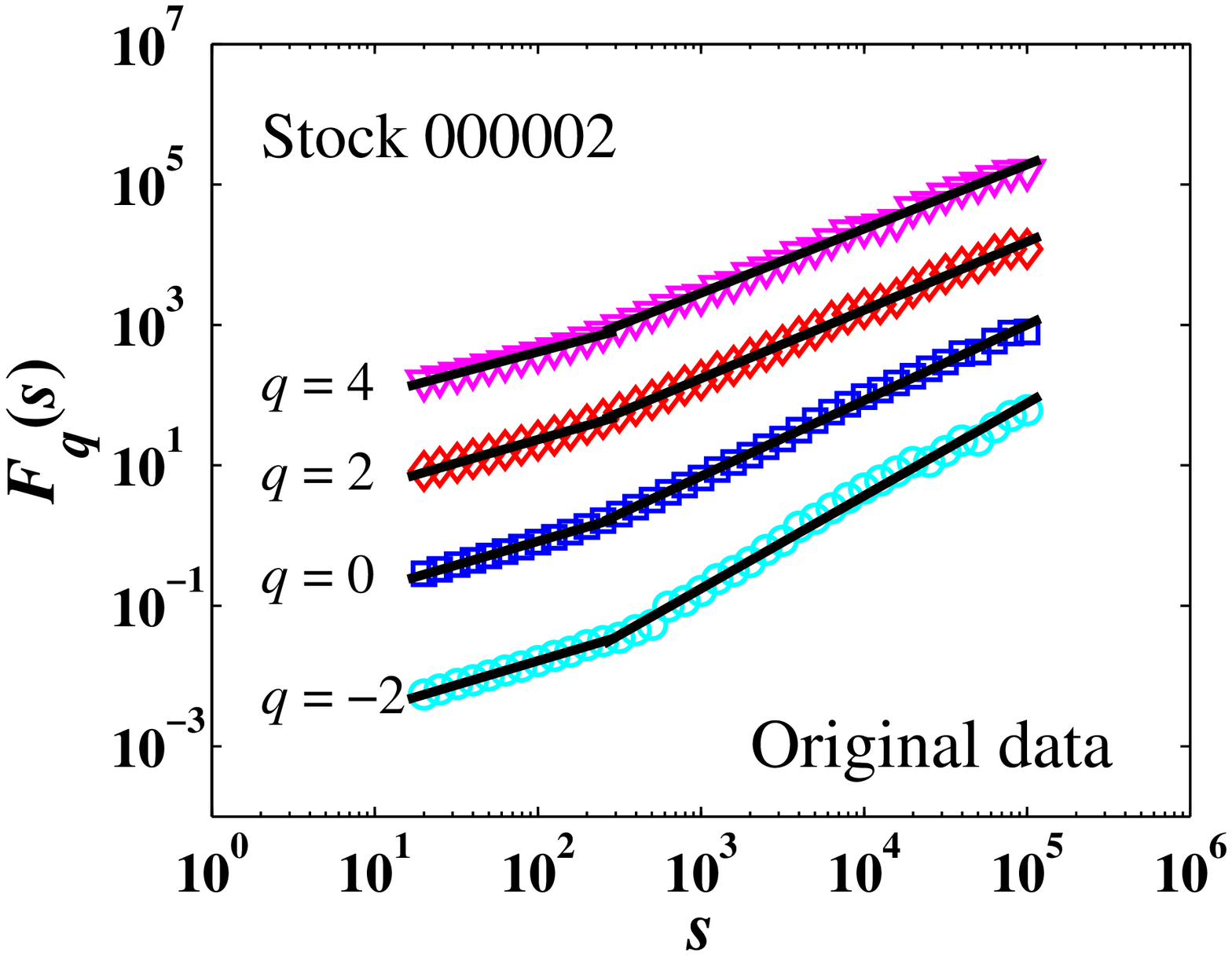}
\includegraphics[width=3.3cm]{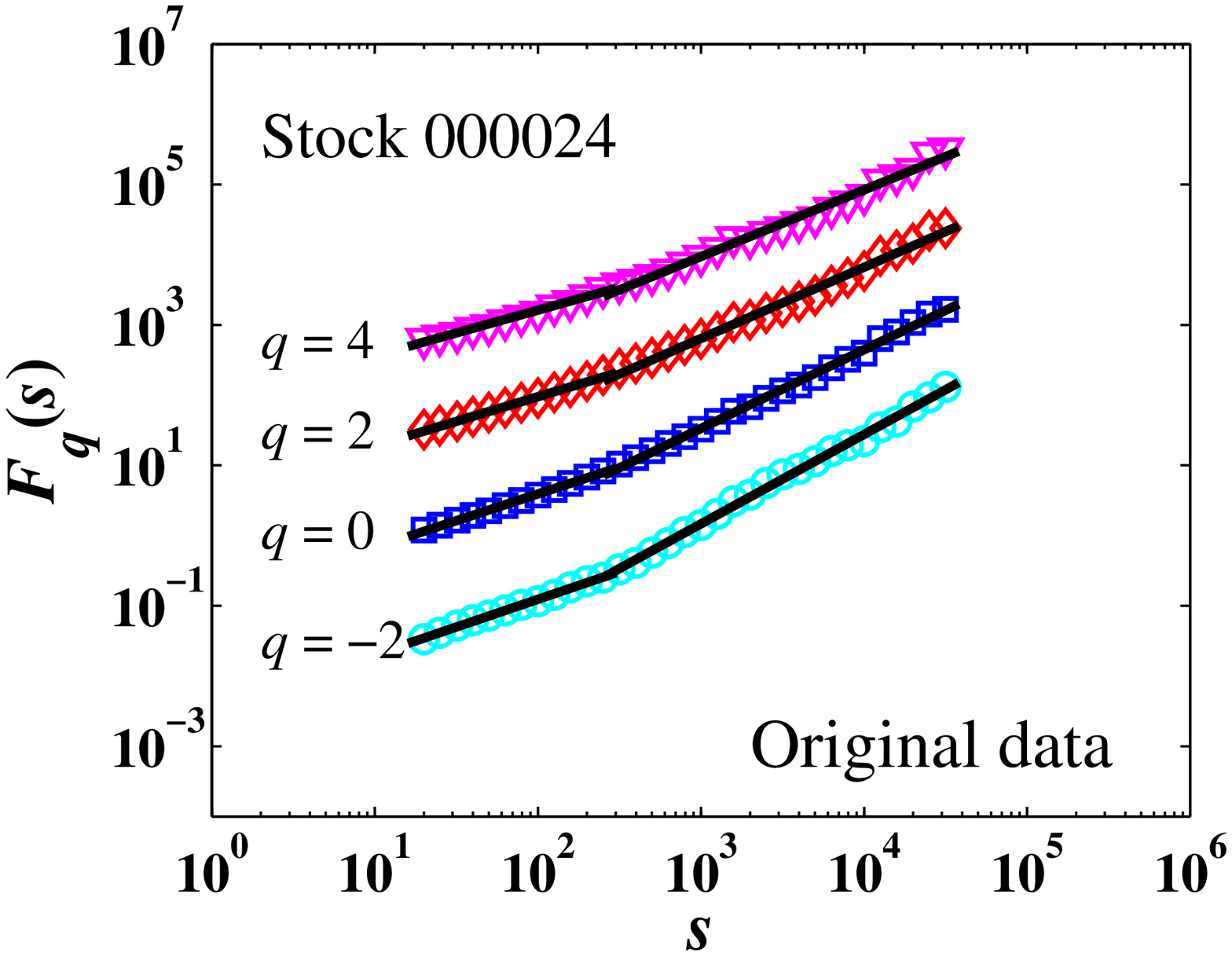}
\includegraphics[width=3.3cm]{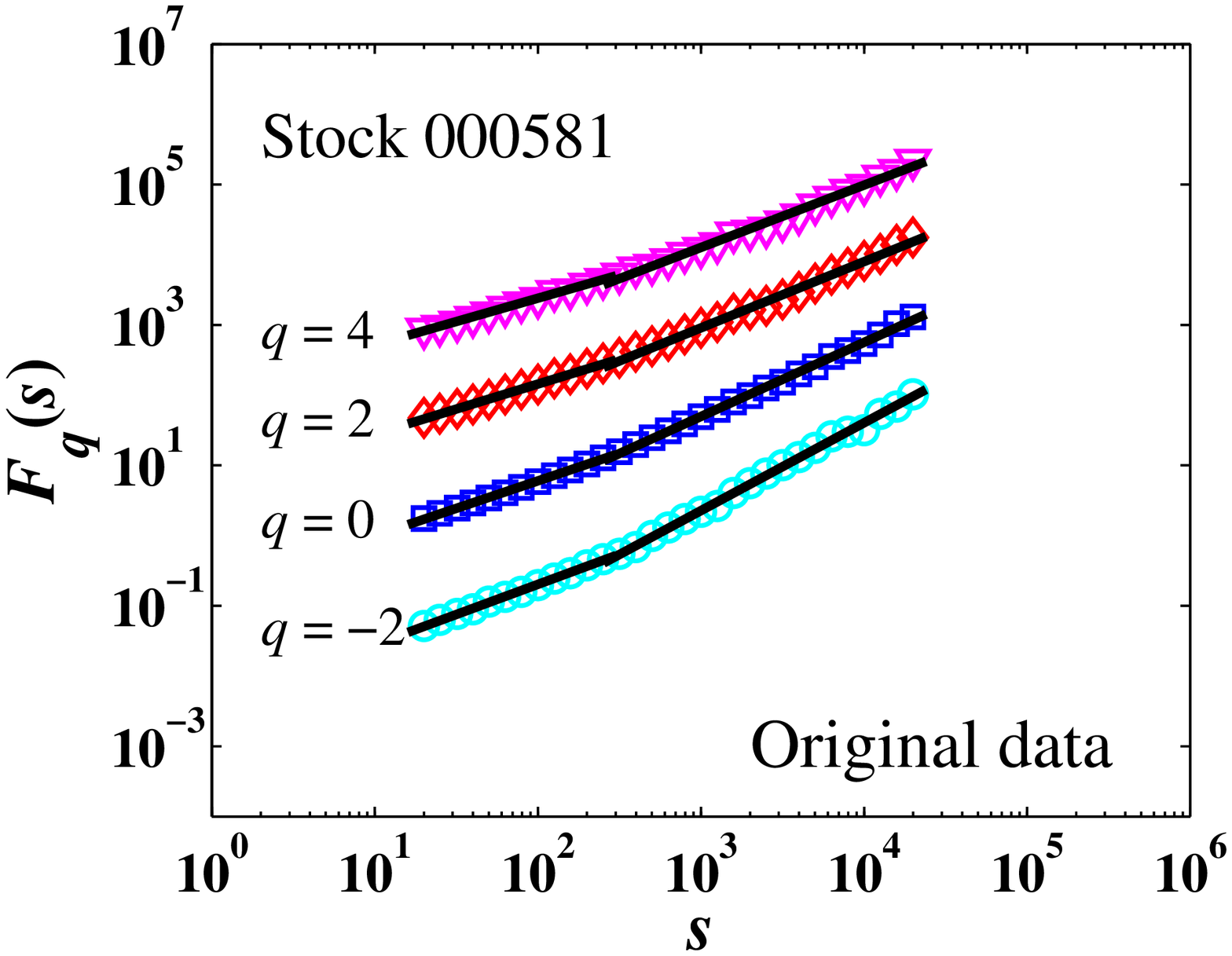}
\includegraphics[width=3.3cm]{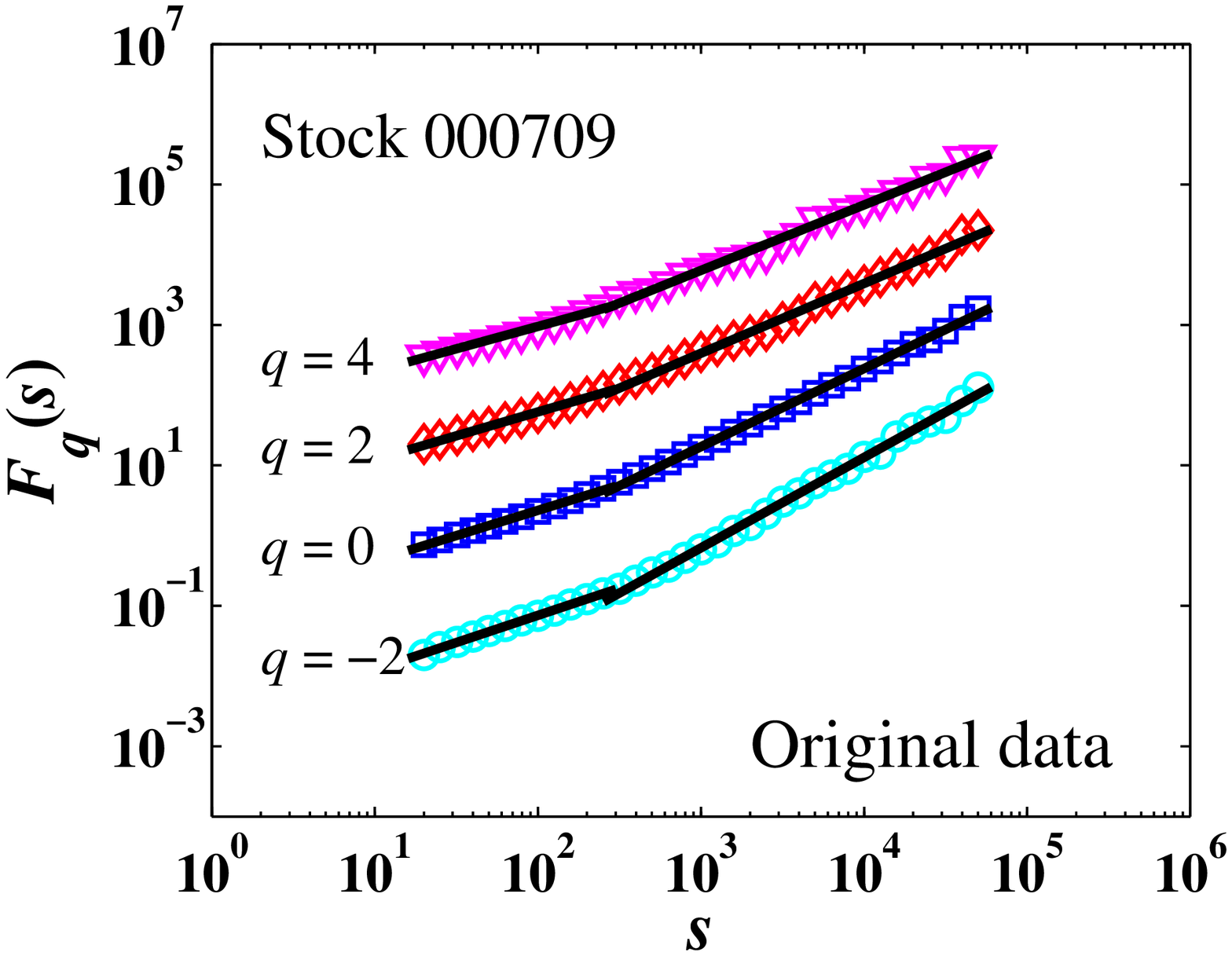}
\includegraphics[width=3.3cm]{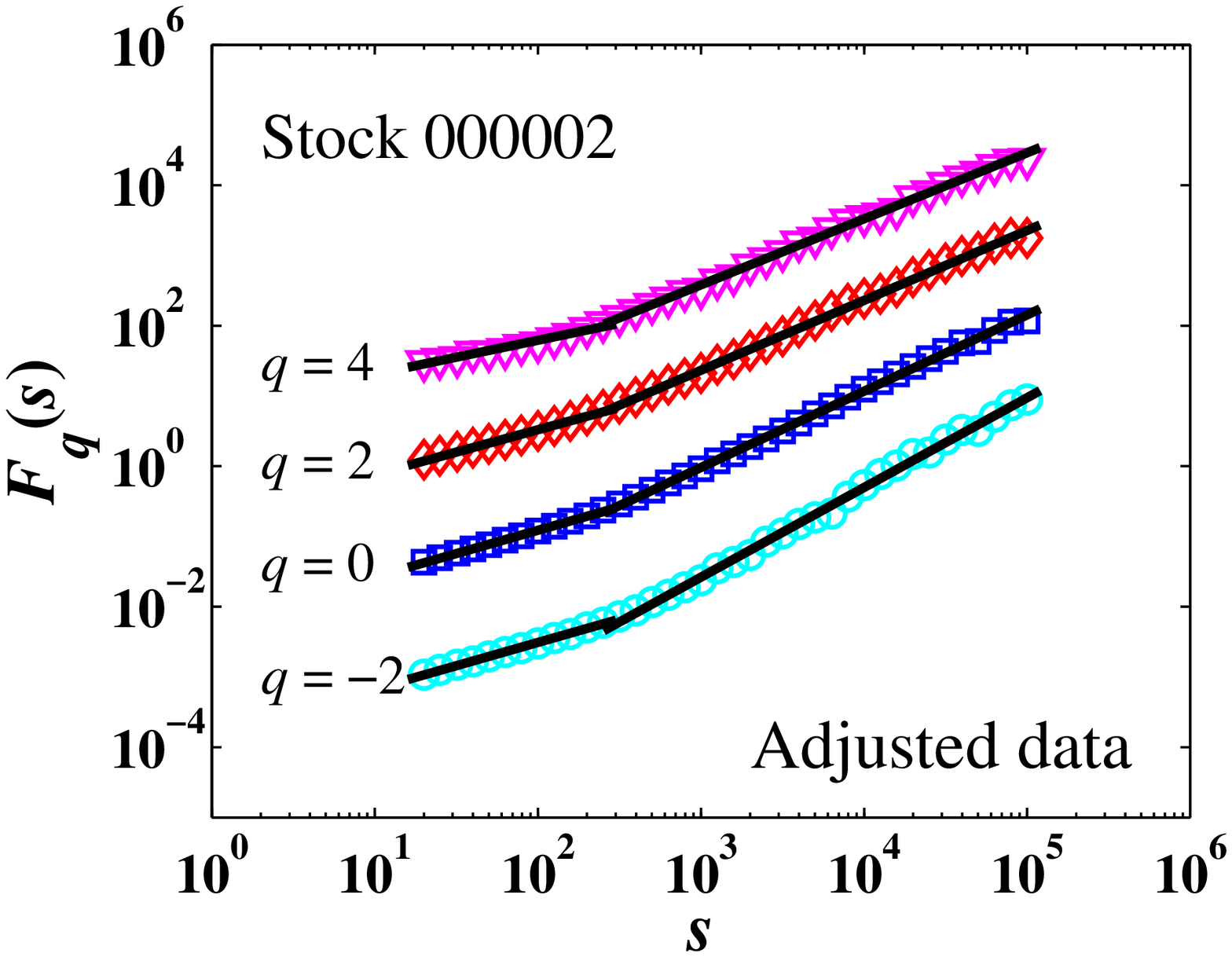}
\includegraphics[width=3.3cm]{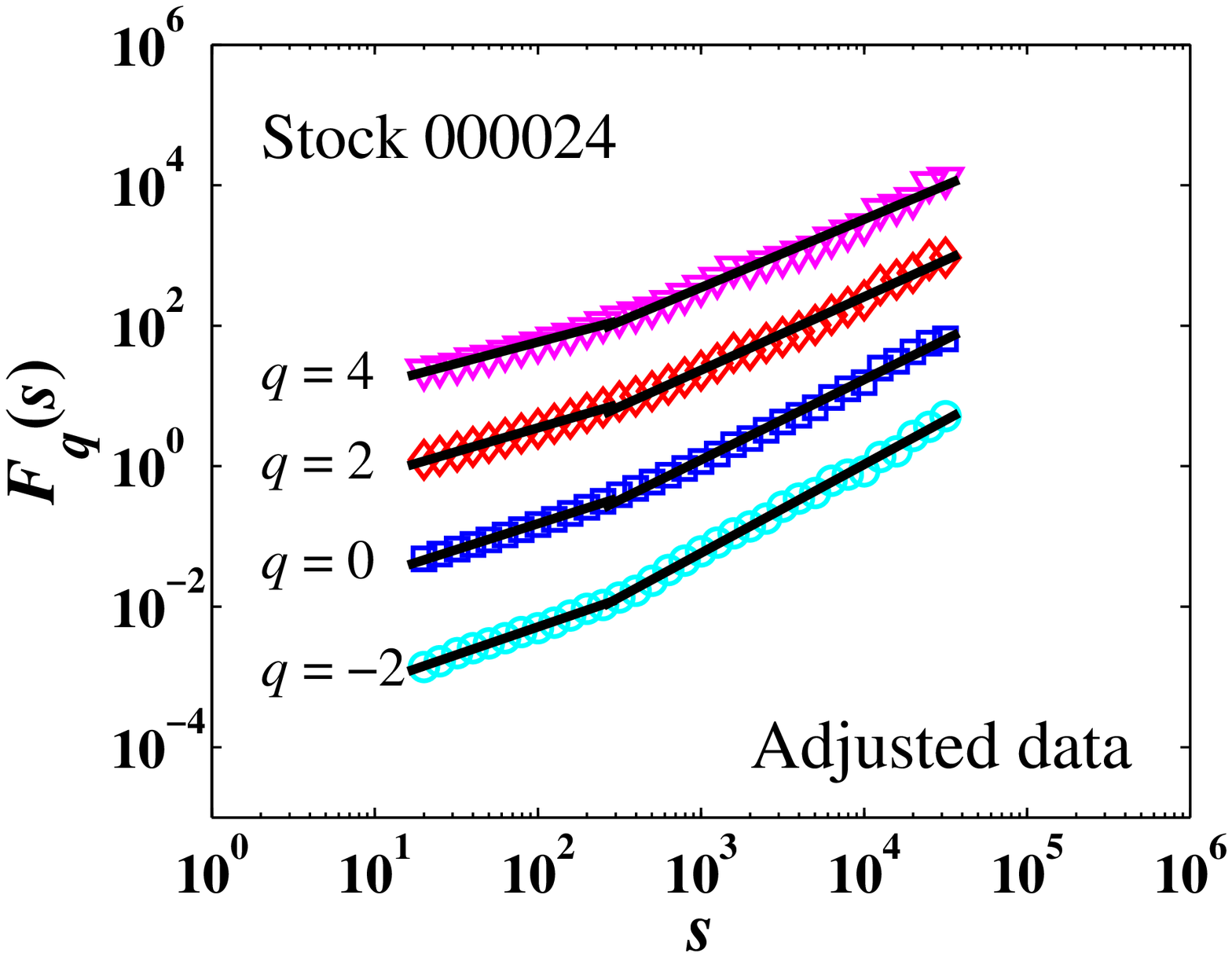}
\includegraphics[width=3.3cm]{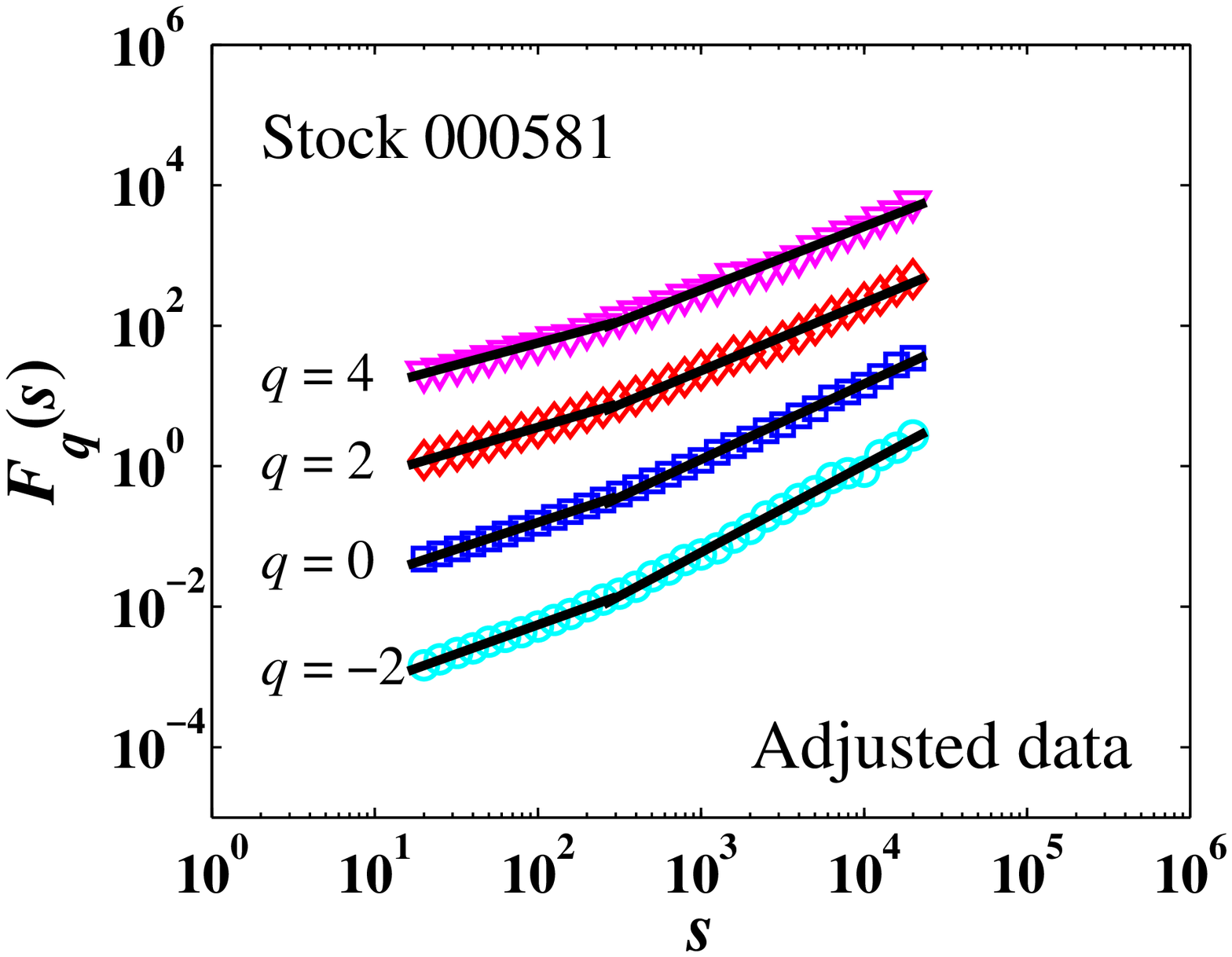}
\includegraphics[width=3.3cm]{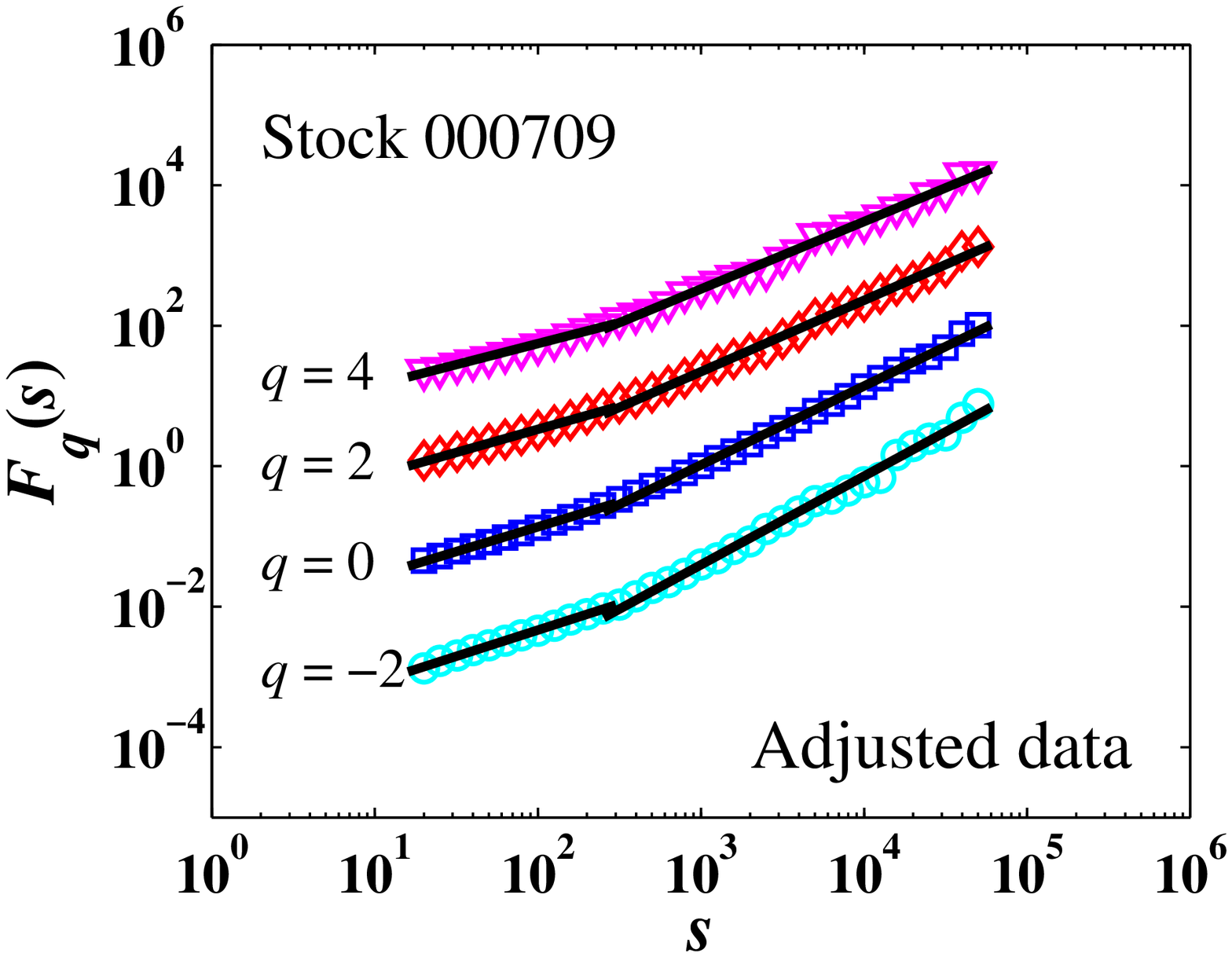}
\caption{(Color online) Plots of the overall fluctuation function
$F_q(s)$ with respect to the box size $s$ for $q = -2$, 0, 2 and 4
in log-log coordinates. The upper and lower panels correspond to the
original data and the adjusted data after removing the intraday
patterns, respectively. The solid lines are the best fits to the
data in the corresponding scaling ranges.} \label{Fig:DFAScaling}
\end{figure}

We present the results of the four typical stocks. For comparison,
the raw data and the adjusted data after removing the intraday
patterns are analyzed. Fig.~\ref{Fig:DFAScaling} shows the
dependence of the overall fluctuation function $F_q(s)$ on the box
size $s$ for different values of $q$ in log-log coordinates. Nice
power laws are observed between $F_q(s)$ and $s$. For each curve,
there is a clear kink indicating the crossover from a power-law
scaling regime at small sizes to a power-law scaling regime at large
sizes. The crossover phenomenon is very common in the detrended
fluctuation analysis of many other financial and physical
quantities.

\begin{figure}[htb]
\centering
\includegraphics[width=4.5cm]{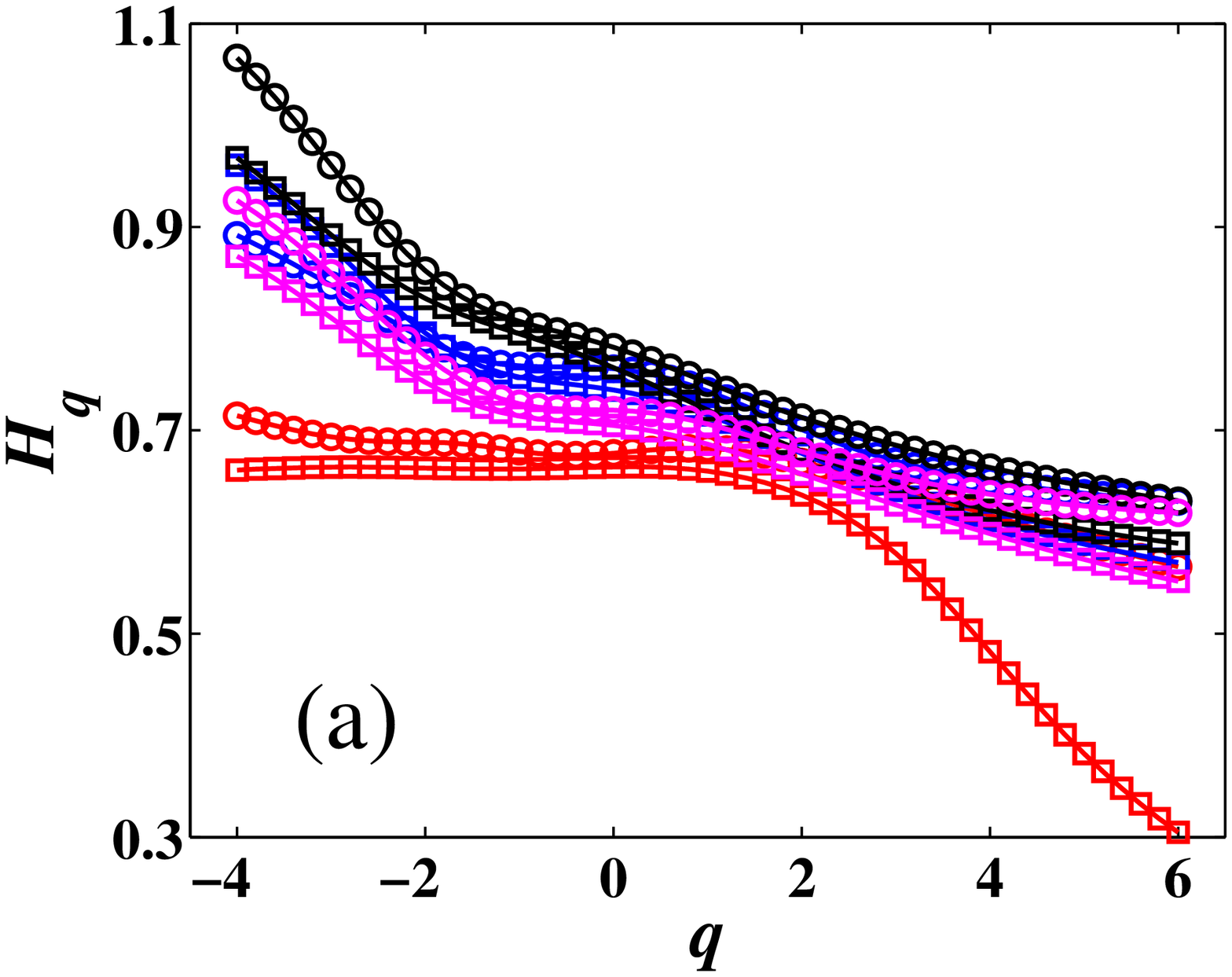}
\includegraphics[width=4.5cm]{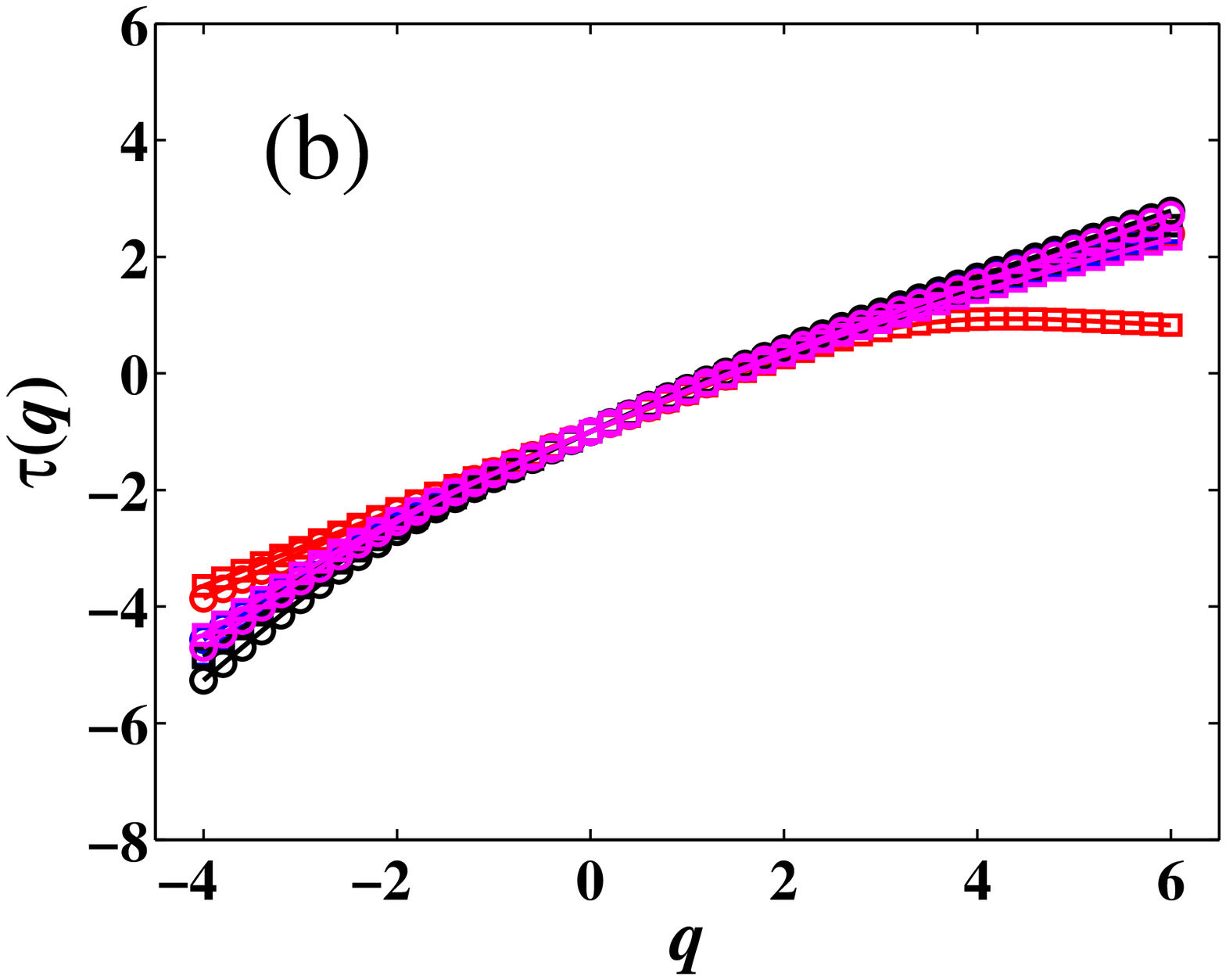}
\includegraphics[width=4.5cm]{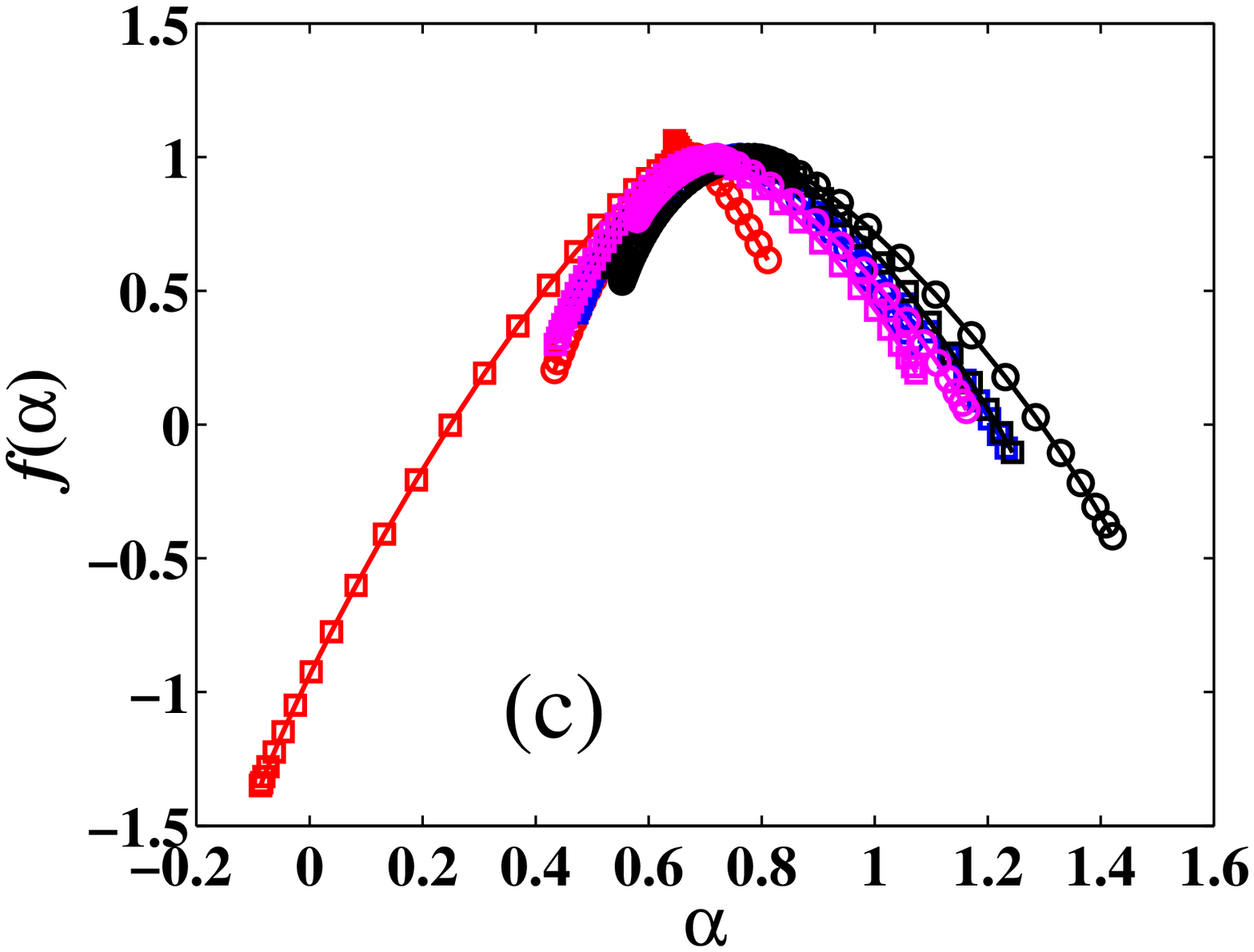}
\includegraphics[width=4.5cm]{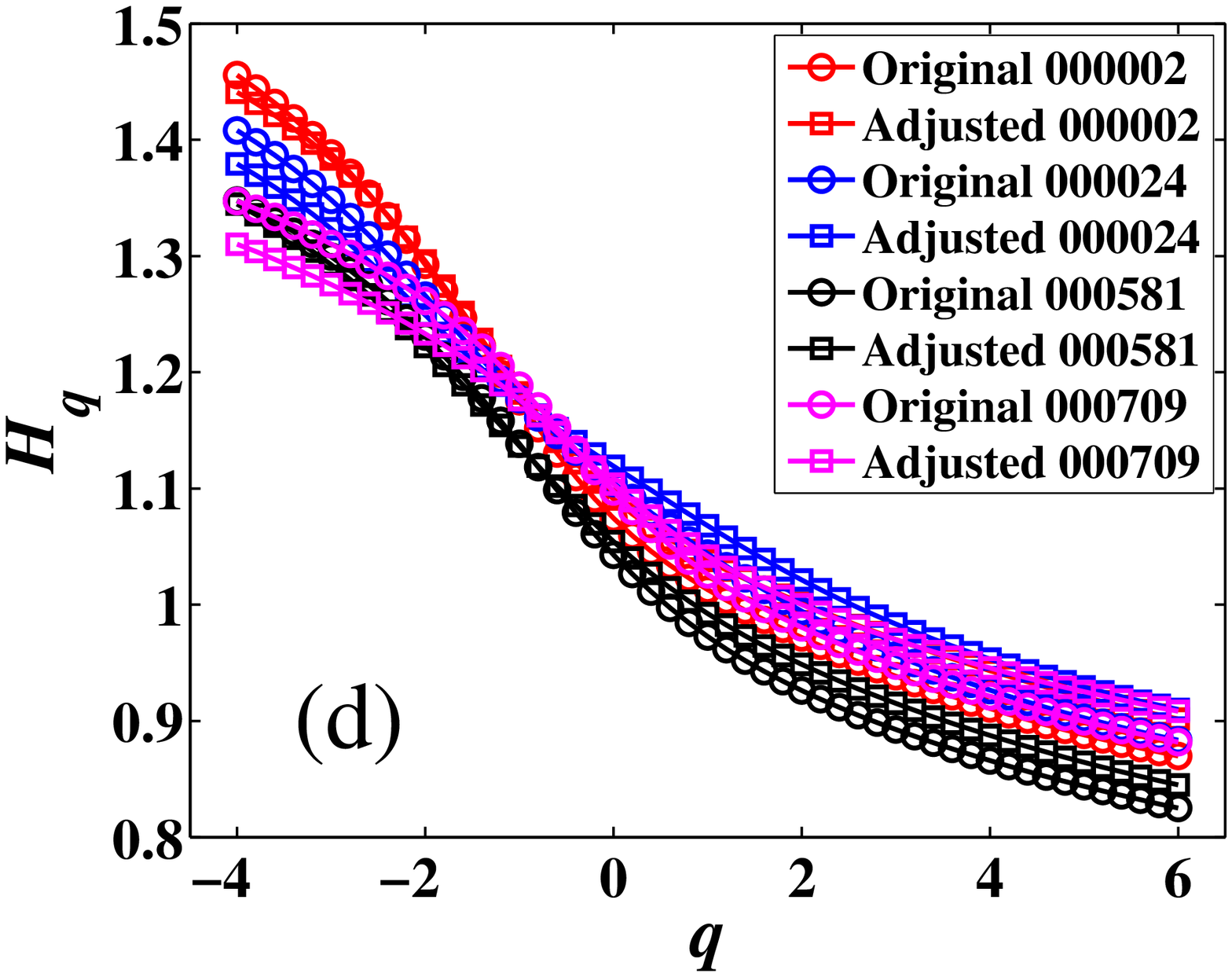}
\includegraphics[width=4.5cm]{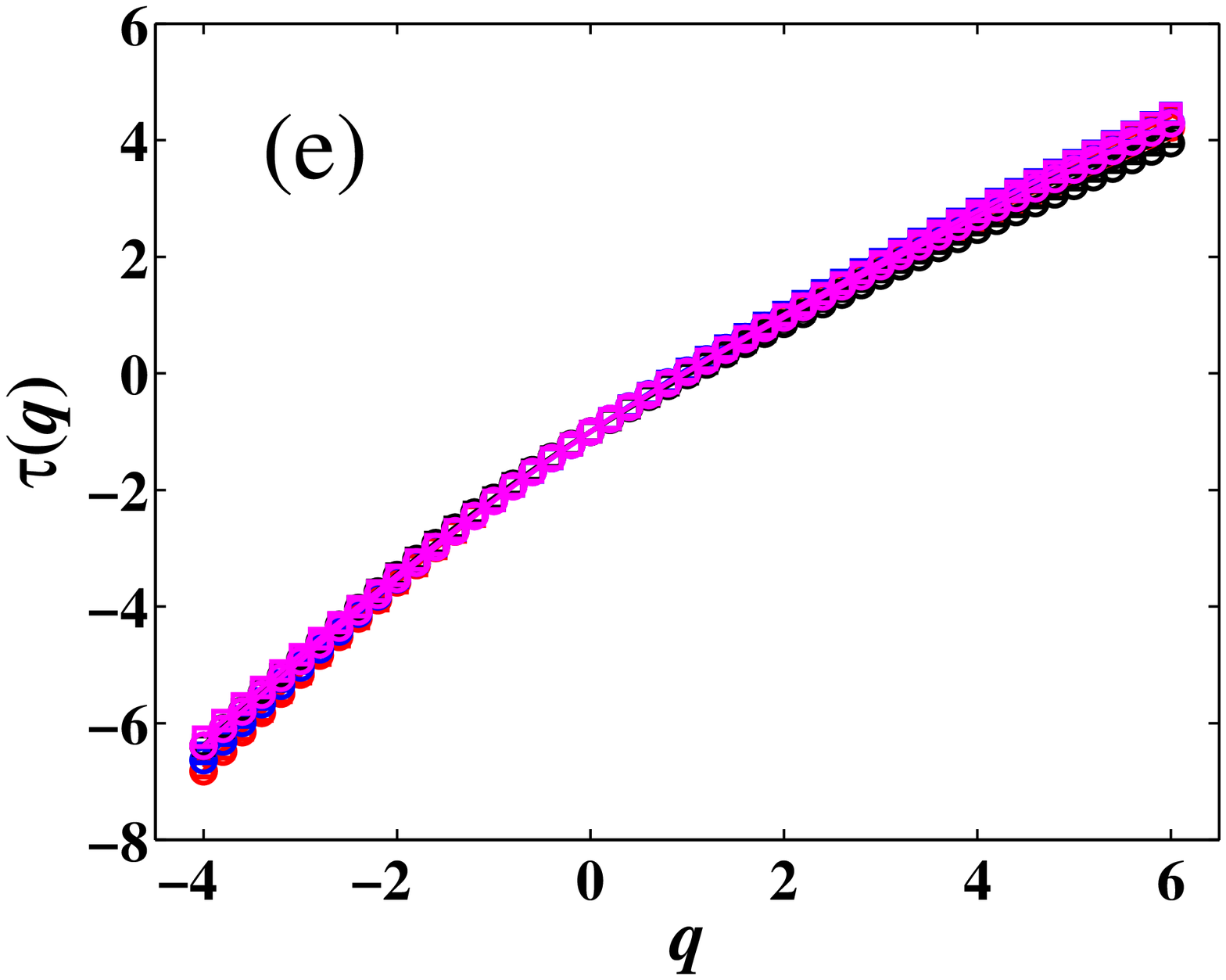}
\includegraphics[width=4.5cm]{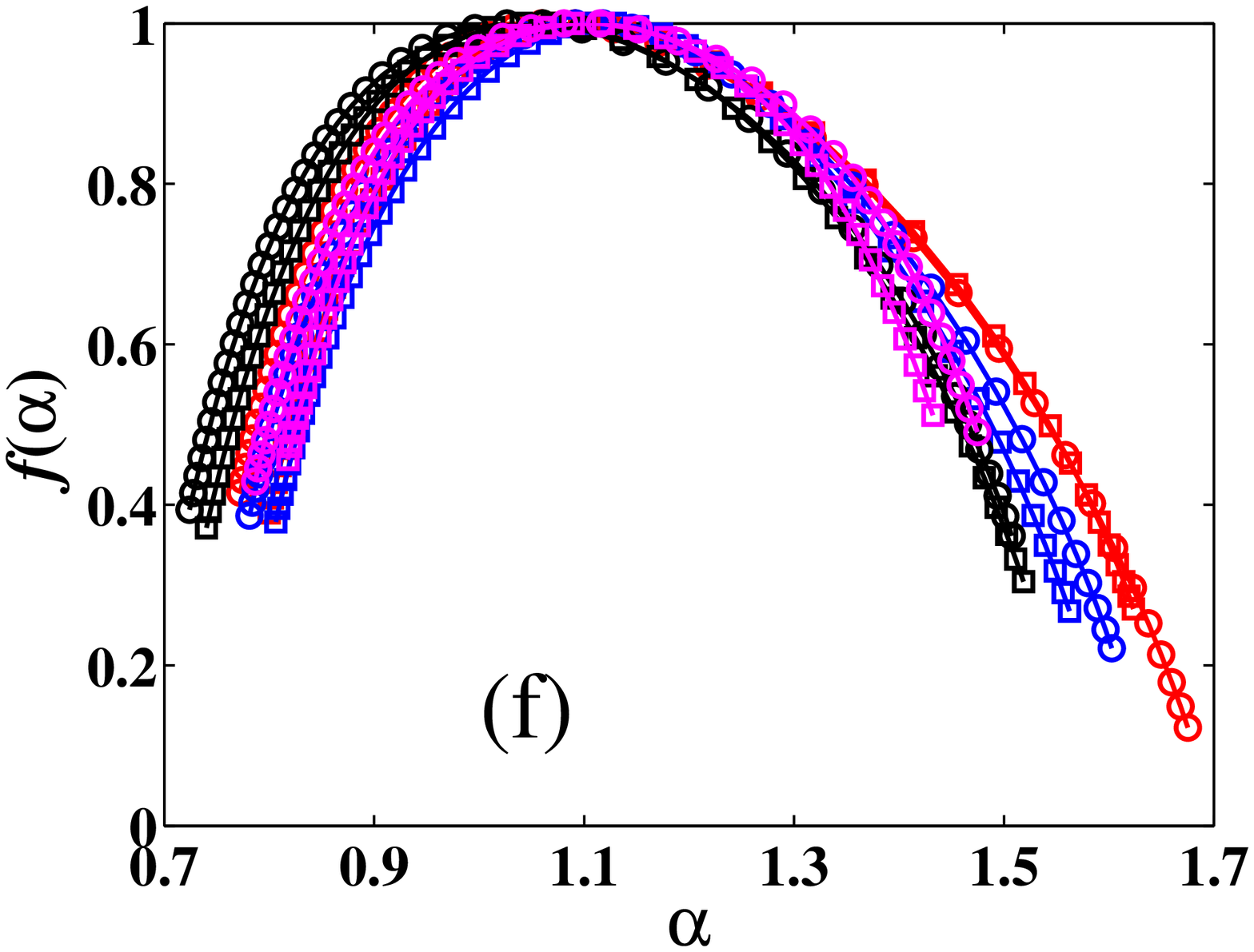}
\caption{(Color online) Multifractal analysis of the intertrade
durations in the small-size regime (upper panel) and the
large-duration panel (lower panel). Shown are the generalized Hurst
indexes $h(q)$ (a, d), the mass exponents $\tau(q)$ (b, e), and the
multifractal spectra $f(\alpha)$ (c, f) for four typical stocks.}
\label{Fig:MF}
\end{figure}

For each case, the scaling exponents in both regimes can be obtained
by the linear regression between $\ln[F_q(s)]$ and $\ln{s}$ in the
two scaling ranges. The estimated generalized Hurst indexes $h(q)$
are illustrated in Fig.~\ref{Fig:MF}(a) for the small-size regime
and in Fig.~\ref{Fig:MF}(d) for the large-size regime. According to
Eq.~(\ref{Eq:scalingfunction}), the mass exponents $\tau(q)$ are
estimated. As shown in Fig.~\ref{Fig:MF}(b) and
Fig.~\ref{Fig:MF}(e), the $\tau(q)$ functions exhibit strong
nonlinearity, which is a hallmark of multifractality.
Fig.~\ref{Fig:MF}(c) and Fig.~\ref{Fig:MF}(f) illustrate the
multifractal singularity spectra $f(\alpha)$, which are obtained by
the Legendre transformation of the mass exponents $\tau(q)$. It is
well-known that $\Delta \alpha \triangleq \alpha_{\max} -
\alpha_{\min}$ is an important parameter qualifying the width of the
extracted multifractal spectrum. The lager the $\Delta \alpha$
value, the stronger the multifractality. The values of the
singularity width $\Delta\alpha$ are listed in
Table~\ref{Tb:HurstIndex} for all the 23 stocks. According to the
lower panel of Fig.~\ref{Fig:MF}, the intertrade durations in
large-size regime show a very neat multifractal nature and the
multifractal behaviors of different stocks are comparable to each
other. The situation for the small-size regime is more complicated,
in which several stocks show different behaviors. For instance, the
multifractal spectrum of stock 000002 in Fig.~\ref{Fig:MF}(c) has a
knot around $q=0$. For other stocks, the curves are normal. The
weakness of the multifractal nature of some stocks are not
surprising due to the narrow scaling ranges at small box sizes.

\section{Conclusion}
\label{S1:Conclusion}

We have investigated the intraday pattern, long memory, and
multifractaliy of the intertrade durations using the limit order
book data and order flows of 23 liquid Chinese stocks traded on the
SZSE in 2003. The intertrade duration shows an inverse $U$-shaped
intraday pattern for all the 23 stocks, which means much smaller
intertrade durations and higher trading activities in the open and
close of a trading day. During the noon closure of the market, new
information arrives and the average intertrade duration within the
first minute of the afternoon trading is also significantly smaller.

The original data and the adjusted data after removing the intraday
pattern from the original data are analyzed with the DFA approach.
Except stock 000720, all the other stocks exhibit a crossover
between two power-law scaling regimes. Intertrade durations in the
small-size regime have a relatively small Hurst index, while those
in the large-size regime have a relatively large Hurst index. In
both regimes, the Hurst indexes are evidently greater than 0.5,
confirming long-range memory in the intertrade durations. We also
find that the intraday pattern has little influence on the
long-range dependence.

In addition, the multifractality in intertrade durations is studied
for the two regimes based on the multifractal DFA method. The
large-duration regime exhibits a sound multifractal feature for all
stocks. In contrast, most of the stocks show multifractality in the
small-duration regime. The scaling range of the small-duration
regime is narrow and thus the resultant multifractal properties are
more or less sensitive to the determination of the crossover point.
These results imply that the trading activities are intermittent.

\bigskip
{\textbf{Acknowledgments:}}

This work was partly supported by the National Natural Science
Foundation of China (Grant Nos. 70501011 and 70502007), the Fok Ying
Tong Education Foundation (Grant No. 101086), and the Program for
New Century Excellent Talents in University (Grant No.
NCET-07-0288).

\bibliography{E:/Papers/Auxiliary/Bibliography}

\end{document}